\documentclass[aps,twocolumn,raggedbottom,prb,showpacs,nobalancelastpage,amssymb,groupedaddress,floatfix]{revtex4}
\usepackage{simplewick}
\usepackage{graphicx}
\usepackage{overpic}
\usepackage{amsthm}
\usepackage{amsfonts}
\usepackage{amsmath}
\usepackage{amssymb}
\usepackage{color}
\usepackage{verbatim}
\renewcommand{\hat}[1]{#1}
\newcommand{\ket}[1]{| #1 \rangle}
\newcommand{\bra}[1]{\langle #1 |}
\newcommand{\eff}{\mathsf{eff}}
\newcommand{\sg}{G}
\newcommand{\FQME}{BR}
\newcommand{\GME}{GME}
\newcommand{\RT}{RT}
\newcommand{\Tmat}{TM}
\newcommand{\Kkernel}{time evolution kernel}
\newcommand{\vq}{q}
\newcommand{\contourorpropagator}{contour}

\newcommand{\hPi}{{\text{\Large{$\pi$}}}}
\newcommand{\tunneling}{\mathsf{T}}
\newcommand{\Lop}{\hat{\mathcal{L}}^{I}_\tunneling}
\renewcommand{\u}[2]{ e^{-\rmi \left(\hat{\mathcal{L}}+\hat{\mathcal{L}}_\leads\right)( #1- #2 )}} 
\renewcommand{\P}{\hat{\mathcal{P}}}
\newcommand{\Q}{\hat{\mathcal{Q}}}
\newcommand{\tauz}{\tau}
\newcommand{\sigl}{\sigma}
\renewcommand{\Re}{\mathsf{Re}}

\newcommand{\hrho}{\hat{\rho}}
\newcommand{\leads}{\mathsf{R}}

\newcommand{\tot}{\mathsf{tot}}
\newcommand{\bias}{\mathsf{b}}
\newcommand{\gate}{\mathsf{g}}
\newcommand{\Sct}[1]{\mbox{Sec.\ \ref{#1}}}
\newcommand{\Scts}[2]{\mbox{Secs.\ \ref{#1}}\ and\ \ref{#2}}
\newcommand{\Eq}[1]{\mbox{Eq.\ (\ref{#1})}}
\newcommand{\eq}[1]{(\ref{#1})}
\newcommand{\noEq}[2]{\mbox{#1\ (\ref{#2})}}
\newcommand{\Eqs}[2]{\mbox{Eqs.\ (\ref{#1})}\ and\ (\ref{#2})}

\newcommand{\Fig}[1]{\mbox{Fig.\ \ref{#1}}}
\newcommand{\Figs}[2]{\mbox{Figs.\ \ref{#1},} \ref{#2}}
\newcommand{\App}[1]{\mbox{App.\ \ref{#1}}}
\newcommand{\rmi}{i}
\newcommand{\rme}{{e}}
\newcommand{\rmd}{\mathrm{d}}

\newcommand{\Tp}{{{T}}^{+}}
\newcommand{\Tm}{{{T}}^{-}}
\newcommand{\Tr}{\mathsf{Tr}\,}
\newcommand{\Trover}[1]{\mathsf{Tr}_{\,#1}\,}
\newcommand{\timeint}[4]{ \underset{#1 > #2 > #3 > #4}{\int \rmd #2 \rmd #3 } }
\newcommand{\timeintthree}[5]{ \underset{#1 > #2 > #3 > #4 > #5}{\int \rmd #2 \rmd #3 \rmd #4 } }

\begin{document}
\title{
Density-operator approaches to transport through interacting quantum dots:
simplifications in fourth order perturbation theory
}
\author{S.\, Koller and M.\, Grifoni}
\affiliation{Institut f\"{u}r Theoretische Physik, Universit\"at
Regensburg, 93035 Regensburg, Germany}
\author{M. \, Leijnse}
\affiliation{Nano-Science Center, Niels Bohr Institute,
University of Copenhagen, Universitetsparken 5, 2100~Copenhagen \O ,
Denmark}
\author{M. R. \, Wegewijs}
\affiliation{Institut f\"{u}r Theoretische Physik A, RWTH Aachen - 52056 Aachen, Germany
\\Institut f\"{u}r Festk\"{o}rper-Forschung - Theorie 3, Forschungszentrum J\"{u}lich - 52425 J\"{u}lich, Germany\\JARA - Fundamentals of Future Information Technology}

\date{\today}
\begin{abstract}
Various theoretical methods address transport effects in quantum dots beyond single-electron tunneling,
while accounting for the strong interactions in such systems.
In this paper we report a detailed comparison between three prominent approaches to quantum transport:
the fourth order Bloch-Redfield quantum master equation (\FQME), the real-time diagrammatic technique (\RT)
and the scattering rate approach based on the T-matrix (\Tmat).
Central to the \FQME{} and \RT{} is the generalized master equation for the reduced density matrix.
We demonstrate the exact equivalence of these two techniques.
By accounting for coherences (non-diagonal elements of the density matrix) between non-secular states, 
we show how contributions to the transport kernels can be grouped in a physically meaningful way.
This not only significantly reduces the numerical cost of evaluating the kernels, but also
yields expressions similar to those obtained in the \Tmat{} approach, allowing for a detailed comparison.
However, in the \Tmat{} approach an ad-hoc regularization procedure is required to cure spurious divergences in 
the expressions for the transition rates in the stationary (zero-frequency) limit.
We show that these problems derive from incomplete cancellation of reducible
contributions and do not occur in the \FQME{} and \RT{} techniques, resulting in well-behaved
expressions in the latter two cases.
Additionally, we show that a standard regularization procedure of the \Tmat{} rates employed in the literature 
does \emph{not} correctly reproduce the  \FQME{} and \RT{} expressions. 
All the results apply to general quantum dot models and we present explicit rules for the simplified calculation of the zero-frequency kernels.
Although we focus on fourth order perturbation theory only, the results and implications generalize to higher orders.
We illustrate our findings for the single impurity Anderson model with finite Coulomb interaction in a magnetic field.
\end{abstract}
 \maketitle

\section{Introduction\label{intro}}
The experimental progress in fabrication of ultrasmall electrical devices~\cite{Tarucha96,Tans97,Martel98,Tans98,Collier99,Bachtold01,Huang01} has made quantum dots one of the standard components in fundamental research and application oriented nanostructures.
Whereas high-resolution transport measurements in the low temperature regime have reached a high degree of sophistication and reveal data dominated by complex many-body phenomena~\cite{DeFranceschi01,Liang02,Park02,vanderWiel03rev,Yu04c60,Schleser05,Jarillo-Herrero05,Sapmaz06,Osorio07a,Parks07,Hauptmann08},
theoretical methods are still struggling to describe these, mainly due to competing influences of strong local interactions and quantum fluctuations~\cite{SW66,Anderson70a,Anderson70b,Koenig97,Koenig98,Paaske04Kondo,Pustilnik04rev,Thorwart05,Schoeller09b,Koch06,Holm08,Leijnse08a,Leijnse09b,Begemann10,Andergassen10}.
\begin{figure}[h!]
\includegraphics[width=0.9\columnwidth]{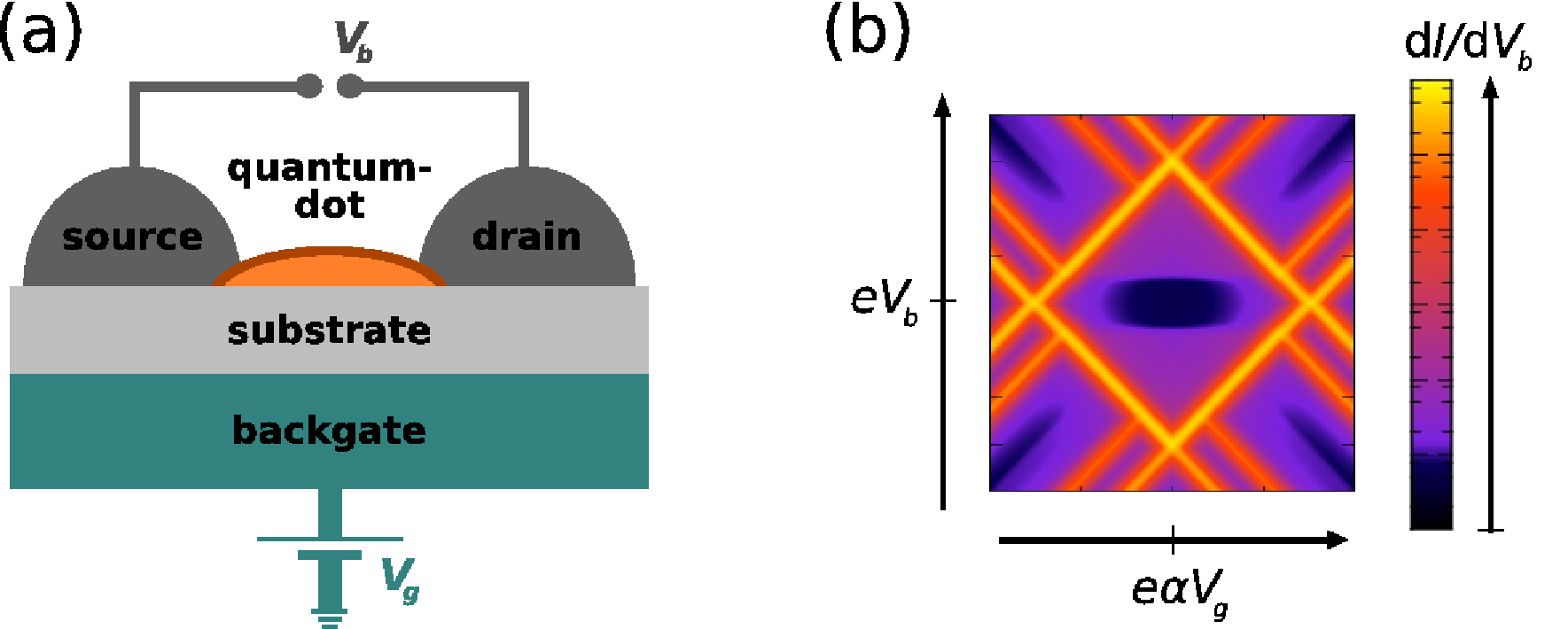}
\caption{(a) Typical transport measurement circuit setup. The source-drain bias 
voltage $V_\bias$ drives a current $I$ through the 
quantum dot. Applying a voltage $V_\gate$ to the gate electrode (in this case a backgate beneath an insulating substrate) 
shifts the energy required to add an additional electron to the dot by $-e \alpha V_\gate$, 
where $-e$ is the electron charge and $\alpha$ the gate coupling.
(b) The measured differential conductance is usually presented in the form of a stability diagram, i.e., in a gate voltage - bias voltage plane. Here we show the characteristics for a single interacting, spin-split level, i.e., an Anderson model in magnetic field, which we return to in \Sct{group} and \Sct{T-mat}
(see \Fig{SD-dIdV} for details).\label{QD}}
\end{figure} 
A common setup for transport studies is drawn in \Fig{QD}(a): the number of electrons on the device is controlled capacitively via a gate voltage $V_\gate$, a difference in the electro-chemical potentials of the leads is created by a bias voltage $V_\bias$.
The measured quantity is the current $I$ or the differential conductance $\rmd I / \rmd V_\bias$ of the whole circuit, which is usually represented in stability diagrams, where the changes in current vs. $V_\gate$ and $V_\bias$ are color-coded [\Fig{QD}(b)].
The observed phenomena strongly depend on the strength of the coupling of the nanodevice to the electronic reservoirs.
In the limit of extremely weak coupling, current at low bias is completely blocked in wide ranges of the gate voltage, showing up as so-called Coulomb diamonds [e.g. \Fig{QD}(b), central region].
Outside these regions of Coulomb blockade only single electrons can be transferred sequentially onto or out of 
the dot~\cite{IngoldNazarov92,Kouwenhoven97rev}, a process called single-electron tunneling (SET).
For simple systems (e.g. without orbital degeneracies) in this regime rate equations~\cite{Kreuzer}
are the standard technique to calculate the occupations of the dot states, the current and other transport quantities~\cite{Bruus}.
The transition rates are calculated by Fermi's Golden Rule, i.e., leading order perturbation theory in the tunneling.
For more complex quantum dots with degenerate orbitals~\cite{Wunsch05,Harbola06,Mayrhofer06,Donarini06,Begemann08,Donarini09,Schultz09,Schultz10} and/or non-collinear magnetic electrodes\cite{Braun04set,Weymann07a,Koller07,Hornberger08}, \emph{coherences}, 
i.e., non-diagonal density matrix elements, give crucial contributions to the 
transport quantities and cannot be neglected.
These are typical situations in molecular electronics and spintronics~\cite{Reckermann09a}.\\
Since the transparency of the contacts is a matter of the material choice as well as of fortune, on the way to low ohmic contacts, intermediate coupling strengths are often observed, allowing for coherent tunneling of multiple electrons~\cite{Averin92}.
Also, there is the possibility to design structures with tunable tunnel barriers, such that different coupling regimes can be systematically accessed, allowing for more detailed spectroscopic information to be extracted~\cite{DeFranceschi01}.
Therefore, electron transport theory must go beyond lowest order perturbation theory in the tunneling, 
while including many charge states, their complex excitations and their quantum coherence.
In recent years, several advanced approaches that address higher-order effects have been developed based on iterative real-time path-integral methods~\cite{Weiss08}, scattering-states~\cite{Mehta06} combined with quantum-monte Carlo~\cite{Han07} or numerical~\cite{Bulla08rev,Anders08} or analytical renormalization group methods~\cite{Paaske04Kondo,Schoeller09b}.
Although these new methods are promising, the standard generalized master equation (GME) approach 
still offers several advantages.
The GME describes the reduced density matrix of the quantum dot with transport kernels that are calculated perturbatively.
The GME can be derived using various methods~\cite{Timm08}: the Nakajima-Zwanzig~\cite{Nakajima58,Zwanzig60}
projection operator technique~\cite{Breuer,Fick,Kuhne78}, the real-time diagrammatic technique~\cite{Schoeller94,Koenig97,Schoeller97hab} (\RT) and the Bloch-Redfield approach~\cite{Wangsness53,Bloch57,Redfield65} (\FQME).
Evaluating the kernels up to fourth order in the tunneling Hamiltonian (next-to-leading order), 
one can account for all processes involving coherent tunneling of one or two electrons. 
These corrections to SET can be calculated either analytically for simple models
or, in complex cases, in a numerically efficient way.
In this case the GME is clearly limited to moderate values of the tunnel coupling as compared to temperature.
However, it has the benefit of non-perturbatively treating both the interactions on the dot as well as the non-equilibrium conditions imposed by the 
bias voltage.
It can therefore provide crucial physical insights into measurements of non-linear transport through complex quantum dots,
see e.g. Ref.~\cite{Huettel09,Zyazin10}.
More generally, even higher corrections can be explicitly formulated in any order of the tunneling by systematic diagram rules.
Since the explicit form of the kernel is known in this way,
a renormalization-group theory for its calculation can be formulated as well,
allowing the non-equilibrium low-temperature regime to be addressed~\cite{Schoeller09a}, including the Kondo effect~\cite{Schoeller09b}.
A class of contributions beyond fourth order can also be included by expanding an equation of motion for the density 
matrix~\cite{Nyvold05, Nyvold10}.

\par
This paper focuses on the \FQME{} and \RT{} formulations of the GME approach applied to transport in the fourth order of perturbation theory and addresses crucial technical matters and simplifications relevant for the description of complex quantum dots. Several important concrete issues have motivated this work:
\par
(i) There is an ongoing discussion about the validity and \emph{equivalence} of approaches which can
become obscured by the complexity of the expressions involved when discussing complex quantum dots.
Clearly, the general form of the quantum master equation is well known since several decades.
Still, a much debated issue is the actual task of systematically calculating higher order corrections to 
the transport kernels occurring in this equation for general complex quantum dot models.
This paper emphasizes that the \FQME{} and \RT{} techniques are \emph{one-to-one} equivalent. 
In contrast, the scattering rate approach based on the generalized Fermi's golden
rule
and T-matrix (\Tmat), as formulated in the literature, differs from these two techniques~\cite{Timm08}.
Although it also relies on a fourth order
perturbative calculation, the results do not coincide in general for identical models. The 
reason is that the objects calculated perturbatively, i.e. the T-matrix and the
time evolution kernel, respectively, are different objects whose relation needs to be clarified. In particular, the
divergences occurring in the \Tmat{} method are intrinsic to the method and \emph{not to the problem}. We show
that these go back to a lack of cancellation of divergent, reducible contributions to the transport kernels
and that the regularizations proposed in the literature cannot reproduce the exact fourth order kernel.
We quantitatively demonstrate the resulting deviations from the correct GME (BR or RT) result for the example 
of a single impurity Anderson model in magnetic field
and analytically show how the divergent \Tmat{} expressions are automatically regularized in the GME approaches.
The GME approaches consistently account for all contributions to the perturbation expansion of the transport kernels in a given order.
The importance of this was recently highlighted for the well studied non-equilibrium Anderson model, which was 
found to exhibit a previously unnoticed resonance due to coherent tunneling of electron pairs~\cite{Leijnse09b}.
\par
(ii) The importance of non-diagonal elements in lowest order calculations involving degenerate states
has long been recognized (``secular contributions''), and continues to attract attention in the context of transport.
Only recently, the importance of non-secular terms (coherences between non-degenerate states) was
found to be crucial~\cite{Leijnse08a} for fourth order tunnel effects.
We generalize the discussion in Ref.~\cite{Leijnse08a} and show how these non-secular corrections can 
efficiently be included into effective fourth order transport kernels through certain reducible diagrams.
\par 
(iii) Explicit expressions for the fourth order transport kernels for a very general class of 
quantum dots were derived in Ref.~\cite{Leijnse08a}. However, the numerical cost of evaluating these
expressions limits their applicability to systems where a relatively small number of many-body excitations ($\lesssim 100$)
has to be accounted for.
Here we show how contributions to the effective kernels can be grouped, making generally 
valid cancellations explicit and resulting in fewer and simpler terms in the perturbation expansion. 
From direct comparison between numerical implementations of the expressions in Ref.~\cite{Leijnse08a} 
and of our new "grouped" expressions, we find the latter to be between 10 and 20 times faster,
\emph{without introducing any additional approximation}.
This allows the treatment of more complicated and realistic quantum dot models.
The direct gain due to the reduction of the number of expressions amounts to a decrease of the computation time by a factor of 4.
However, the grouping structure can be exploited further to make the numerical implementation more efficient, leading to the additional speed-up.
Moreover, the grouping gives a basis upon which an explicit connection to \Tmat{} expressions can be revealed.
\par
As will become clear in the course of this paper, our newly found grouping intimately connects and enlightens the above three issues, which warrants
our systematic and extended discussion.
The key ideas presented can be applied to analyze higher order contributions as well as to similar perturbation and renormalization group calculations for other classes of problems.
\par
The structure of the paper is as follows.
In \Sct{modelham} we discuss the model Hamiltonian of the setup \Fig{QD}(a) and some pertinent notation.
We then introduce the reduced density matrix (RDM) describing the quantum dot as part of the whole system
and the generalized master (or kinetic) equation (\GME{}) which describes its time-evolution. We summarize its
general properties and the common ground of the discussed approaches.
We then turn to the derivation of the generalized master equation using the \FQME{} and \RT{} technique.
The crucial role played by time-ordering, irreducibility of contributions and analytic properties
(lack of spurious divergences) is emphasized.
The derivations are given as compactly as possible, because there exists 
a broad formal study on different master-equation approaches by Timm~\cite{Timm08}, 
who also showed their equivalence. 
In contrast to
his work, our comparison continues in \Sct{diagr-repr} on a more explicit level with a mapping between
terms arising from the \FQME{} and \RT{} diagrams.
In \Sct{nonsec} we demonstrate how the non-secular contributions of coherences lead to important corrections
in the fourth order transport rates. 
Based on this, \Sct{group} introduces a grouping of contributions to the transport kernels, yielding significant 
simplifications due to partial cancellations, which is followed by an analysis of how the groups of diagrams contribute to fourth order physical transport processes (cotunneling, pair tunneling, and level renormalization and broadening).
In \Sct{T-mat} the derivation of the \Tmat{} is reformulated.
The long-standing problem of a precise comparison with the \RT{} technique in the context of transport 
theory and the origin of divergences in the \Tmat{} approach is solved for the general case.
The theoretical discussion in the last two sections, \ref{group} and \ref{T-mat},
is illustrated by the tangible application to a single impurity Anderson model in magnetic field.
 
\section{Model and generalized master equation\label{modelham}}
The standard model for a quantum dot system coupled to contacts reads
\begin{equation}
  \hat{H}_\tot=\hat{H}+\hat{H}_\tunneling+\hat{H}_\leads\,.\label{totham}
\end{equation}
The Hamiltonian 
\begin{equation}\label{tunham}
\hat{H}_\leads=\sum_{l=s,d}\sum_{\sigl}\sum_{\vq}\left(\epsilon_{l\sigma \vq}-\mu_{l}\right)\hat{c}^\dag_{l\sigl\vq}\hat{c}_{l\sigl\vq}
\end{equation}
models the reservoirs, i.e., the source and the drain contact.
The operator $\hat{c}_{l\sigl\vq}^\dag$ ($\hat{c}_{l\sigl\vq}$) creates (annihilates) an electron in a 
state $\vq$ with energy $\epsilon_{l\sigma \vq}$ in the source ($l=s$) or drain ($l=d$) contact,
where $\sigl$ denotes the spin projection.
The bias voltage shifts the electro-chemical potentials of the source and drain leads such that $\mu_{s}-\mu_{d} = eV_\bias$, where $-e$ is the electron charge. 
\\
The coupling between the quantum dot and the leads is described by the tunnel Hamiltonian
\begin{multline}
  \hat{H}_\tunneling 
=\sum_l\hat{H}_{\tunneling l}\\\equiv
\sum_{l} \sum_{\sigl\vq m}\left(t_{l m \vq}\,
    \hat{d}^{\dag}_{\sigl m}\hat{c}_{l\sigl\vq}+t^*_{l m \vq}\,\hat{c}^{\dag}_{l \sigl\vq}\hat{d}_{\sigl m}\right),
  \label{lt_ham}
\end{multline}
where $\hat{d}^{\dag}_{\sigl m}$ ($\hat{d}_{\sigl m}$) creates (annihilates) an electron in the single particle state $m$ on the dot.
The single-particle amplitude $t_{lmq}$ for tunneling from an orbital state $q$ in lead $l$ to an orbital state $m$ on the dot is assumed to be independent of spin. 
Finally, the dot is described by the Hamiltonian
\begin{equation}
	\hat{H} = \sum_a E_a |a \rangle \langle a |,
\label{dot_ham}
\end{equation}
where $|a\rangle$ is a \emph{many-body} eigenstate of the dot with energy $E_a$.
The precise dependence of these energies on the applied voltages arising from capacitive effects (see e.g., Ref.~\cite{Kouwenhoven97rev, Delft01rev}) is irrelevant for the following discussion. Typically, the gate voltage dependence is linear, $E_a \propto-eV_\gate N_a$, where $N_a$ is the number of electrons for state $a$.

The diagonalized many-body Hamiltonian~(\ref{dot_ham}) together with the \emph{tunnel matrix elements} (TMEs) $T^{\pm}_{l\sigma q}(a,a')$ between all the many-body eigenstates $a,\,a'$,
\begin{subequations}\begin{eqnarray}
T^{+}_{l\sigma q}(a,a')&:=&\sum_m t_{lmq}\left\langle a\left|\hat{d}^{\dag}_{\sigl m}\right| a'\right\rangle,
\label{tme_manybody1}
\\
T^{-}_{l\sigma q}(a,a')&:=&\left[T^{+}_{l\sigma q}(a',a)\right]^{*},
\label{tme_manybody}
\end{eqnarray}\end{subequations}
form the crucial input to the \GME{} transport theory, which 
thereby incorporates local interaction effects non-perturbatively.
Here $T^{\pm}_{l \sigma q}(a,a')$ is only nonzero if the state $a$ differs from $a'$ in its electron number by $N_a-N_{a'}=\pm1$.
Typically, the TMEs can be assumed independent of $q$, but this is not a prerequisite for the results of this work.

We focus on the regime of weak tunnel coupling and thus split $\hat{H}_\tot$ in a free part 
$\hat{H}_0=\hat{H} +\hat{H}_\leads$ and a perturbation $\hat{H}_\tunneling$. 
The condition for weak coupling is that the broadening $\hbar \Gamma_l$ induced by tunneling processes from and to lead $l$ is small compared to the thermal energy, i.e., $\hbar \Gamma_l \ll k_B T$, 
where $\hbar$ and $k_B$ are respectively the Planck and Boltzmann constants. The broadening is defined by
\begin{equation}\label{Gammal}\Gamma_{l}=\frac{2\pi}{\hbar}\sum_{\sigma q}\left|t_{lmq}\right|^2\delta(\epsilon_{l\sigma q}-\omega),
\end{equation}where for the contexts in which we need it here (i.e., as a measure of order of magnitude), both the orbital ($m$) dependence and the frequency ($\omega$) dependence are neglected. Notice that the analytical expressions we derive in this work are a priori \emph{not} subject to these restrictions.
For simplicity, we will further denote contributions of the order of $\Gamma_l$ in general by $\Gamma$. 
In this paper we will go beyond lowest order in $\Gamma$, allowing the regime of intermediate coupling 
to be addressed.
\par
Throughout the paper, we will make use of the Liouville (superoperator) notation, in addition to standard 
Hamiltonian operator expressions.
The former is merely an efficient bookkeeping tool on a general level, whereas the latter may be more useful
when evaluating matrix elements.
For example, for the dot Hamiltonian the abbreviation
\begin{equation}\label{lioudef}
  \hat{\mathcal{L}} \hat{B}:= \frac{1}{\hbar}\left[\hat{H}, \hat{B} \right]
\end{equation}
defines the action of the Liouville superoperator in the Schr\"odinger picture on an arbitrary operator $B$.
It generates the time-evolution through $e^{\rmi \hat{\mathcal{L}}t} \hat{B}= e^{\frac{\rmi}{\hbar} \hat{H}t} \hat{B} e^{-\frac{\rmi}{\hbar} \hat{H}t}=B(t)$
(Baker-Campbell-Hausdorff formula). Analogous expressions hold for the other Hamiltonians, and in particular we will need
\begin{equation}
  \Lop(t) \hat{B}^I(t') = \frac{1}{\hbar}\left[\hat{H}^I_\tunneling(t), \hat{B}^I(t') \right],
\end{equation}
where $\hat{B}^I(t')=e^{\frac{\rmi}{\hbar}\left(\hat{H}+\hat{H}_\leads\right)t'}\hat{B}e^{-\frac{\rmi}{\hbar}\left(\hat{H}+\hat{H}_\leads\right) t'}= e^{\rmi \left(\hat{\mathcal{L}}+\hat{\mathcal{L}}_\leads\right) t'} \hat{B}$ is an operator in the interaction picture. Notice that the time evolution of the superoperator in the interaction picture is thus
\begin{equation}\label{time-lop}
\Lop(t)=e^{\rmi \left(\hat{\mathcal{L}}+\hat{\mathcal{L}}_\leads\right)t} \hat{\mathcal{L}}_\tunneling \,e^{-\rmi\left(\hat{\mathcal{L}}+\hat{\mathcal{L}}_\leads\right) t}.
\end{equation}

\subsection{Generalized master equation and steady state}
\label{easy}
The object of interest here is the {reduced density matrix}~\cite{Blum_book} (RDM)
\begin{equation}
  \hrho(t) = \Trover{\leads}\left\{\hrho_\tot(t)\right\}.
\end{equation} 
It describes the state of the quantum dot incorporating the presence of the leads, which are traced out of the total density matrix $\hrho_\tot$, as prescribed by $\Tr_\leads$.
Once $\hrho(t)$ is known, the expectation value of any observable can be calculated, as discussed below.
When the interaction $\hat{H}_\tunneling$ is switched on at time $t=t_0$, the total density matrix $\hrho_\tot$ is the direct product of the (arbitrary) initial state $\hrho(t_0)$ of the quantum dot and the equilibrium state $\hrho_\leads$ of the leads,
\begin{align}
  \hrho_\leads = \frac{e^{-H_{\leads}/(k_BT)}}{Z_\leads}
  \label{rholeads}
\end{align}
with $Z_\leads=\Trover{\leads} e^{-H_{\leads}/(k_BT)}$.
After the interaction $H_\tunneling$ is switched on, i.e. for times $t>t_0$, correlations, which are of the order of the tunnel coupling~\cite{Blum_book},
build up between leads and quantum dot, causing $\hrho_\tot$ to deviate from the factorized form:
\begin{align}
  \hrho_\tot(t)
   &= e^{-\rmi \hat{\mathcal{L}}_\tot (t-t_0)} \hrho_\tot(t_0)
 \nonumber\\
  &=\hrho(t)\hrho_\leads(t)+\theta(t-t_0)\,\mathcal{O}(\hat{\mathcal{L}}_\tunneling).
  \label{rho-sep1}
\end{align}
We emphasize that it is crucial to include in a kinetic equation for $\hrho(t)$, 
the correlations $\mathcal{O}(\hat{\mathcal{L}}_\tunneling)\simeq\mathcal{O}(\hat{H}_\tunneling)$ between leads and quantum dot
consistently beyond linear order in $\hat{H}_\tunneling$, if one is interested in going beyond lowest order.
As we will see in \Scts{Nak-Zwa}{RTD}, the \RT{} approach incorporates them automatically by directly integrating
out the leads for times $t>t_0$, while for the \FQME{} one explicitly solves for the deviation from the factorized
state. Both the \FQME{} and the \RT{} technique lead to the \textit{generalized quantum master equation} (or \textit{kinetic equation}), describing the time evolution of the RDM 
\begin{equation}\label{rdm-compact}
	\dot{\hrho}(t)=-\rmi \hat{\mathcal{L}} \hrho(t) + \int_{t_0}^t\!\!\rmd \tau\ \hat{\mathcal{K}}(t-\tau)\hrho(\tau).
\end{equation}
Here, the first term accounts for the time evolution due to the local dynamics of the quantum dot.
In the second time non-local term, the \textit{\Kkernel} $\hat{\mathcal{K}}(t - \tau)$ is a superoperator
acting on the density operator.
Convoluted in time with $\hrho(\tau)$, it gives that part of the time evolution which is generated by the tunneling.
We note that this form of the GME is dictated by the linearity of the Liouville equation and the partial
trace operation.
\par
The kernel to fourth order formally reads
\begin{widetext}
  \begin{align}
    \hat{\mathcal{K}}(t-\tau)\hrho(\tau)
    \label{thekernel}
    =&
    - \Trover{\leads}\left\{ \hat{\mathcal{L}}_\tunneling \u{t}{\tau} \hat{\mathcal{L}}_\tunneling \hrho(\tau)\hrho_\leads\right\}
    \\\nonumber
    &+\timeint{t}{\tau_2}{\tau_1}{\tau} 
    \Bigl[
      \Trover{\leads}  \left\{     \hat{\mathcal{L}}_\tunneling \u{t}{\tau_2}     \hat{\mathcal{L}}_\tunneling \u{\tau_2}{\tau_1}
                            \hat{\mathcal{L}}_\tunneling \u{\tau_1}{\tauz} \hat{\mathcal{L}}_\tunneling  \hrho(\tau) \hrho_\leads\right\} \Bigr.\\\nonumber
     &\qquad\Bigl. -
      \Trover{\leads}\left\{\hat{\mathcal{L}}_\tunneling \u{t}{\tau_2}     \hat{\mathcal{L}}_\tunneling e^{-\rmi\left(\hat{\mathcal{L}}+\hat{\mathcal{L}}_\leads\right)\tau_2} \hrho_\leads\
      \Trover{\leads}\left\{e^{\rmi\left(\hat{\mathcal{L}}+\hat{\mathcal{L}}_\leads\right)\tau_1}\hat{\mathcal{L}}_\tunneling \u{\tau_1}{\tauz} \hat{\mathcal{L}}_\tunneling  \hrho(\tau)\hrho_\leads \right\}\right\}
    \Bigr]. 
  \end{align}
\end{widetext}
Below we will show how \FQME{} and \RT{} (as well as the Nakajima-Zwanzig projection operator approach, see \App{NZ}), when consistently applied, lead to this result.
The central topic of this paper is the explicit evaluation of \Eq{thekernel} and the significant simplifications which can be achieved in the steady state limit $\lim_{t\to\infty}\hrho(t)=\hrho$.
In this limit, \Eq{rdm-compact} becomes
\begin{equation}
  \lim_{t\to\infty}\dot{\hrho}(t)
  = 0 =
	-\rmi \mathcal{L} \hrho +\hat{K}\hrho,\label{rho-lapl}
\end{equation}
where $\hat{K} = \hat{K}(z = \rmi0)$ and $i0$ denotes an imaginary infinitesimal, 
and \begin{equation}\label{kern-lapl}\hat{K}(z)=\int_0^\infty \rmd\tau \ \rme^{iz \tau} \, \hat{\mathcal{K}}(\tau)\end{equation} is the Laplace transform of the \Kkernel{}.
Taking matrix elements with respect to the many-body eigenstates of the dot Hamiltonian, $\hat{H}$, 
we obtain from \Eq{rho-lapl} a set of linear coupled equations for all states $a,a'$ of the RDM:
\begin{equation} 
  0 = -\rmi\sum_{aa'} \delta_{ab}\,\delta_{a'b'}\left(E_a-E_{a'}\right)
  \rho_{aa'}+\sum_{aa'}{{K}}^{aa'}_{bb'}\rho_{aa'}.
  \label{vectoreq-1}
\end{equation}
Here, the matrix elements of $\hat{K}$ (or any other superoperator) are defined by
\begin{align}
{{\hat{K}}}^{aa'}_{bb'} := \langle b |\,  [\, \hat{K} | a\left\rangle\right\langle a' |\, ]\, | b' \rangle ,
\label{supermatel}
\end{align}
where we use square brackets to make clear that the kernel superoperator must first act on  
$| a\left\rangle\right\langle a' |$, and then the matrix elements of the resulting operator are taken.
Each diagonal element of the RDM equals the probability of finding the system in a certain state. Thus,
the normalization condition
\begin{equation}
  \sum_{a}\rho_{aa}=1.
  \label{norm}
\end{equation} 
must be fulfilled. The \noEq{restriction}{norm} allows the system of linear equations obtained from \Eq{vectoreq-1} to be solved, since without it they are under-determined due to the sum-rule
\begin{equation}
  \sum_{b}K^{aa'}_{bb}=0\quad\forall\,a,a'.\label{sumrule}
\end{equation} 
Physically, this guarantees that gain and loss of probability are balanced in the stationary state.
\par
The expectation value of any non-local observable can be expressed in a form similar to \Eq{rdm-compact}.
In particular, we can write the particle current flowing out of lead $l$ (i.e. the number of electrons leaving lead 
$l$ per unit time) as
\begin{equation}
  \label{current}
  I_l(t)
  = \left\langle\hat{I}_l(t)\right\rangle
  =  \Tr \int_{t_0}^t\!\!\rmd \tau\ \hat{\mathcal{K}}_{I_l} (t - \tau)\hrho(\tau),
\end{equation}
where $\hat{\mathcal{K}}_{I_l}(t - \tau)$ is the kernel associated with the current operator
\begin{align}
  \hat{I}_l
  &= -\frac{\rmi}{\hbar} \left[\hat{H}_{Tl},\hat{N}_l\right]
  \nonumber
  \\
  &= -\frac{\rmi}{\hbar} \sum_{\sigma q m} t_{l m q} \hat{d}_{\sigma m}^{\dag}\hat{c}_{l \sigma q} + \text{h.c.}\ ,
  \label{currentoperator}
\end{align}
with $\hat{N}_l=\sum_{\sigl{q}}\hat{c}^\dag_{l\sigl{q}}\hat{c}_{l\sigl{q}}$ being the number operator in lead $l$.
\\
Taking the steady state limit of \Eq{current}, the stationary current is given by the zero-frequency component 
$\hat{K}_{I_l}:=\hat{K}_{I_l}(z = i0)$ of the Laplace transform of the current kernel, traced over in product with 
the stationary density matrix $\hrho$:
\begin{equation}
  I_l=\mathsf{Tr}\left\{\hat{K}_{I_l}\hrho\right\}=\sum_{b}(\hat{K}_{I_l})^{aa'}_{bb}\rho_{aa'}.
  \label{DC-current}
\end{equation}
The current kernel can be calculated by simple modification of the time-evolution kernel as discussed below in subsections \ref{Nak-Zwa} and \ref{RTD} explicitly.
We will now address the derivation of \Eq{rdm-compact} and of its kernel \eq{thekernel} 
up to fourth order in the tunnel coupling.
We focus here on two approaches, an iterative procedure in the time
domain~\cite{Wangsness53,Bloch57,Redfield65}, referred to as Bloch-Redfield approach
(\FQME{}), and the RT approach~\cite{Schoeller94,Koenig97,Schoeller97hab} 
(\RT{}).
The projection operator technique of Nakajima\cite{Nakajima58} and Zwanzig \cite{Zwanzig60},
which has been explained and used in many works~\cite{Breuer,Fick,Kuhne78,Jang01},
is closely related and equivalent to the \FQME{} approach 
and is discussed for completeness in \App{NZ}.
\\
The derivation of the kinetic equation requires no other
ingredient than the Liouville equation for the total density matrix $\hrho_\tot$~\cite{Blum_book}:
\begin{equation}
  \dot{\hrho}_\tot^{I}(t)=-\rmi\,\Lop(t)\hrho_\tot^{I}(t).
  \label{liouville-new}
\end{equation}
For the purposes of this paper, it is most convenient to work in the time-domain and use the interaction picture.
In addition, we make use of the property of the particular bi-linear coupling of \Eq{lt_ham} considered here,
that the lead-average of an odd number of interactions vanishes due to the odd number of lead field operators
in $\hat{H}_\tunneling$.

\subsection{Bloch-Redfield approach\label{Nak-Zwa}}
The Bloch-Redfield approach~\cite{Wangsness53,Bloch57,Redfield65} is usually favored to derive 
the second order quantum master equation\cite{Blum_book}.
Basically, one integrates \Eq{liouville-new} and reinserts it back into its differential form to get
\begin{align}
  \dot{\hrho}_\tot^{I}(t)
  =-\rmi\,\Lop(t)\hrho_\tot^I(t_0)
  &-\int_{t_0}^t\rmd\tauz\ \Lop(t)\Lop(\tauz)\hrho_\tot^{I}(\tauz) \label{SQME_diff}.
\end{align}
We now extend this to fourth order~\cite{Laird90} by repeating the iteration steps: 
we transform \Eq{SQME_diff} to an integral equation,
\begin{multline}
  {\hrho}_\tot^{I}(t)=\hrho_\tot^I(t_0)-\rmi\!\int_{t_0}^t\!\rmd\tauz\ \Lop(\tauz)\hrho_\tot^I(t_0)\\
  - \timeint{t}{\tau_1}{\tau}{t_0}
  \Lop(\tau_1)\Lop(\tauz)\hrho_\tot^{I}(\tauz), \label{SQME_int}
\end{multline}
which is once more reinserted into \Eq{liouville-new}.
After integration one arrives at
\begin{multline}
  \hrho_\tot^{I}(t)=
  \hrho_\tot^I(t_0)
  -\rmi \int_{t_0}^t \!\!\rmd\tauz\ \Lop(\tauz)\hrho_\tot^I(t_0)
  \\
  - \timeint{t}{\tau_1}{\tau}{t_0}
  \Lop(\tau_1)\Lop(\tauz)\hrho_\tot^{I}(t_0)
  \\
  + \rmi \timeintthree{t}{\tau_2}{\tau_1}{\tau}{t_0}
  \Lop(\tau_2)\Lop(\tau_1)\Lop(\tauz)\hrho_\tot^{I}(\tauz).
  \label{TQME_int}
\end{multline}
We reinsert \Eq{TQME_int} back into the Liouville \noEq{equation}{liouville-new} and perform the trace over the leads in order to obtain the RDM. Thereby, terms involving in total an odd number of lead operators vanish. Due to the relations $\hrho_\tot^{I}(t_0)=\hrho^I(t_0)\hrho_{\leads}$ and with \Eq{rho-sep1}
we obtain:
\begin{multline}
  \dot{\hrho}^{I}(t)
  =
  -\int_{t_0}^t\!\rmd \tau_2
  \Trover{\leads}\left\{
  \Lop(t)\Lop(\tau_2)\hrho^{I}(t_0) \hrho_\leads\right\}\\
  + \timeintthree{t}{\tau_2}{\tau_1}{\tau}{t_0}
  \Trover{\leads}\left\{
  \Lop(t)\Lop(\tau_2)\Lop(\tau_1)\Lop(\tauz)\hrho^{I}(\tauz)\hrho_\leads\right\}\\
  +\mathcal{O}((\Lop)^6).
  \label{\FQME_prev}
\end{multline}
The second order contribution in \Eq{\FQME_prev} contains $\hrho^{I}(t_0)$ instead of $\hrho^{I}(\tauz)$
and lacks the convoluted form which the fourth order term has:
thus in the stationary limit the initial state $\hrho^{I}(t_0)$ does not seem to drop out.
If one naively were to neglect this difference and set $\hrho^{I}(t_0) \approx \hrho^{I}(\tau)$
the fourth order kernel would contain spurious divergences (see~\Sct{T-mat}).
Instead, one has to account for the correlations between dot and reservoirs at times $t>t_0$
up to order $(\Lop)^2$ as expressed by \Eq{SQME_int}.
Taking the trace over the leads this equation gives:
\begin{align}
  {\hrho}^{I}(\tau_2)
  &=
  \hrho^{I}(t_0)
  - \timeint{\tau_2}{\tau_1}{\tau}{t_0}
  \Trover{\leads} \left\{ \Lop(\tau_1)\Lop(\tauz)\hrho^{I}(\tauz)\hrho_\leads  \right\}
  \nonumber\\
  &+\mathcal{O}((\Lop)^4).
  \label{for_tnl}
\end{align}
This shows also that \Eq{\FQME_prev} through $\hrho^{I}(t_0)$ still contains higher-order terms.
Consistently neglecting these in \Eq{\FQME_prev}, we can eliminate the dependence on the initial 
condition from \Eq{\FQME_prev} and thus arrive at the GME in the interaction picture
\begin{equation}\label{\FQME-compact}
  \dot{\hrho}^I(t)=\int_{t_0}^t\!\!\rmd\tauz\ \hat{\mathcal{K}}^I(t-\tauz)\hrho^I(\tauz),
\end{equation}
with the time-evolution kernel defined by
\begin{widetext}
  \begin{align}
    \hat{\mathcal{K}}^{I}(t-\tau)\hrho^I(\tau)
    &=
    - \Trover{\leads}\left\{ \Lop(t)\Lop(\tauz)  \hrho_\leads \hrho^I(\tau)\right\} +
    \nonumber\\
    \timeint{t}{\tau_2}{\tau_1}{\tau}
    &
    \left[
      \Trover{\leads} \left\{ \Lop(t)\Lop(\tau_2)\Lop(\tau_1)\Lop(\tauz)  \hrho_\leads  \hrho^I(\tau)\right\}
      -
      \Trover{\leads}\left\{\Lop(t)\Lop(\tau_2)\,\hrho_\leads
      \Trover{\leads}\left\{\Lop(\tau_1)\Lop(\tauz)   \hrho_\leads\hrho^I(\tau)\right\}\right\}
    \right]. 
    \label{\FQME-explicitkernel}
  \end{align}
\end{widetext}
Transforming the RDM to the Schr\"odinger picture by  
\begin{equation}\label{ia-to-schr}
\dot{\hrho}^I(t)=e^{\rmi\left(\hat{\mathcal{L}}+\hat{\mathcal{L}}_\leads\right)t}\left(\dot{\hrho}(t)+\rmi\hat{\mathcal{L}}\hrho(t)\right),\end{equation}
and with the Liouville operators according to \Eq{time-lop},
we obtain the generalized master \noEq{equation}{rdm-compact}, 
and arrive at the \noEq{expression}{thekernel} for the kernel.\\

The current, \Eq{current}, is given by\begin{subequations}
\begin{align}
  I_l(t)
  &= \Trover{\tot} \left\{\hat{I}^I_l(t)\hrho^I_\tot(t)\right\}
  \label{current-ia}\\
  &=  \Tr \int_{t_0}^t\!\!\rmd t'\ \hat{\mathcal{K}}^I_{I_l} (t-t') \hrho^I(t'),
  \label{current-ia2}
\end{align}\end{subequations}
where the current kernel in the interaction picture is given by \Eq{\FQME-explicitkernel} with $\Lop(t)$
replaced by $\hat{I}_l(t)$.
In deriving this, as for the density matrix, one must take care to keep the time-ordered structure:
since the current operator $\hat{I}_l(t)$ is, as $\hat{H}_{\tunneling l}$, linear in the lead operators [cf. 
\Eq{currentoperator}], we obtain \Eq{current-ia2}, correct up to fourth order, by inserting
the third order iteration for $\hrho_\tot$ (\Eq{TQME_int}) into \Eq{current-ia}.
Under the trace, the first and third contribution are zero since they contain an odd number of interactions.
Next, as before, in the second contribution of \Eq{TQME_int}, $\hrho(t_0)$ has to be eliminated
using \Eq{for_tnl}, thereby generating a fourth order correction term.
Finally, in the fourth term one must consistently keep
$\hrho_\tot(\tau) \approx \hrho(\tau) \hrho_\leads$, i.e. only here one can neglect the deviation from the
factorized form.

\subsection{Real-time diagrammatic technique\label{RTD}}
The real-time approach has been discussed on a general level in many works~\cite{Schoeller94,Koenig97,Schoeller97hab}.
Therefore the aim of this section is to recall how one efficiently arrives at the kinetic equation and its kernel by exploiting Wick's theorem from the outset.
We start from the Liouville \noEq{equation}{liouville-new} for the full system in the interaction picture and formally integrate it:
\begin{equation} 
\hrho^{I}_\tot (t)=   \hat{\mathcal{T}} e^{-\rmi \int_{t_0}^{t} \rmd \tau \Lop(\tau)} \hrho^{I}_\tot(t_0),
\end{equation}
where $\hat{\mathcal{T}}$ is the time-ordering superoperator.
Using  
$\hrho^{I}_\tot(t_0)
= \hrho_\tot(t_0)
 =\hrho^{I}_\leads(t_0)\hrho^{I}(t_0),
$ and defining
the superoperator
\begin{align}\label{supopPi}
  \hPi^{I}(t,t_0) =  \Trover{\leads}\left\{ \hat{\mathcal{T}} e^{-\rmi \int_{t_0}^{t} \rmd \tau \Lop(\tau)} \hrho_\leads(t_0)\right\},
\end{align} 
the time-evolution of the reduced density matrix can formally be written as:
\begin{align}
  \hrho^{I}(t)=  \hPi^{I}(t,t_0) \hrho^{I}(t_0).
\end{align}

Expanding the time-ordered exponential superoperator, the trace in \Eq{supopPi} can be explicitly evaluated term by term by Wick's theorem:
the trace over each string of reservoir field operators becomes a product of pair contractions, indicated in the following by contraction lines.
For our purposes here, one can simply formally consider the Liouvillians to be contracted (meaning their reservoir part), see \Eq{rt-tnl} below.
We can then decompose $\hPi^{I}(t,t_0)$ into a reducible and an irreducible part, depending on whether or not the contractions 
separate into disconnected blocks.
Collecting all irreducible parts into the kernel $\hat{\mathcal{K}}^I(t-t')$ one obtains in the standard way a Dyson equation: \begin{align}
  \hPi^{I}(t,t_0) =    \hat{1} +
  \timeint{t}{\tau_2}{\tau_1}{t_0}
  \hat{\mathcal{K}}^{I}(\tau_2-\tau_1) \hPi^{I}(\tau_1,t_0).
\label{pi-rt}
\end{align}
It relates the full propagator $\hPi^{I}(t,t_0)$ to the free propagator, which equals unity in the interaction picture, and to the irreducible kernel $\hat{\mathcal{K}}^{I}$.
Applying the Dyson equation to $\hrho(t_0)$ and taking the time derivative, one arrives at the kinetic equation in the interaction picture, \Eq{\FQME-compact}. Transformed back to the Schr\"odinger picture we obtain the kinetic \noEq{equation}{rdm-compact}.
\\
We have now obtained the kernel $\hat{\mathcal{K}}^I(t-\tau)$ formally as the sum of irreducible contributions to the time-evolution superoperators of different orders in the tunneling, which are written down to fourth order:
\begin{align}\nonumber 
  & \hat{\mathcal{K}}^{I}(t-\tau) = -
  \contraction{}{\Lop(t)}{}{\Lop(\tau)}
  \Lop(t)\Lop(\tau)
  +
  \timeint{t}{\tau_2}{\tau_1}{\tau}
  \\
  \Bigl[
  &\label{rt-tnl}
  \contraction    {}{\Lop(t)}{\Lop(\tau_2)}{\Lop(\tau_1)}
  \contraction[2ex] {\Lop(t)}{\Lop(\tau_2)}{\Lop(\tau_1)}{\Lop(\tau)}
                     \Lop(t)  \Lop(\tau_2)  \Lop(\tau_1)  \Lop(\tau)
  +
  \contraction       {\Lop(t)}{\Lop(\tau_2)}{}{\Lop(\tau_1)}
  \contraction[2ex]{}{\Lop(t)}{\Lop(\tau_2)    \Lop(\tau_1)}{\Lop(\tau)}
                      \Lop(t)  \Lop(\tau_2)    \Lop(\tau_1)  \Lop(\tau) \Bigr],
\end{align}
where $
  \contraction{}{\Lop(t)}{}{\Lop(\tau)}
                 \Lop(t)    \Lop(\tau)
  := \Trover{\leads} \left\{\Lop(t)\Lop(\tau) \hrho_\leads\right\}
$.

The expectation value of the current (or of any operator) is obtained in a similar fashion.
We first introduce a superoperator $\hat{\mathcal{L}}_{I_l}^I=\frac{1}{2}\{\hat{I}_l^I, \cdot \}$, 
which is an anti-commutator in contrast to the superoperators governing the time-evolution.
The current is then expressed as
\begin{align}
  I_l^I(t) =  \Tr\left\{\hPi^{I}_{I_l}(t,t_0) \hrho^{I}(t_0)\right\} ,
\end{align}
where we introduced a current propagator
\begin{align} 
  \hPi^{I}_{I_l}(t,t_0) =  \Trover{\leads} \left\{\hat{\mathcal{T}} \hat{\mathcal{L}}_{I_l}^I(t) e^{-\rmi \int_{t_0}^{t} \rmd \tau \hat{\mathcal{L}}^I(\tau)} \hrho_\leads(t_0) \right\},
\end{align}
which differs from the propagator for the reduced density operator only by the current vertex $\hat{\mathcal{L}}_{I_l}^I(t)$
at the final time. 
Collecting all parts of $\hPi^{I}_{I_l}$ which are irreducibly connected to the latter vertex,
one readily verifies that the remaining irreducible parts at earlier times are those contained in the propagator $\hPi^{I}$:
\begin{align}
  \hPi^{I}_{I_l}(t,t_0) = 
  \timeint{t}{\tau_2}{\tau_1}{t_0}
  \hat{\mathcal{K}}^{I}_{I_l}(\tau_2 - \tau_1) \hPi^{I}(\tau_1,t_0).
\end{align}
The current kernel $\hat{\mathcal{K}}^I_{I_l}(t - \tau)$ is given formally as the sum of irreducible contributions
to the time-evolution superoperators of different orders in the tunneling with the leftmost interaction vertex 
replaced by the current superoperator.
Applying this equation to $\hrho(t_0)$ and taking the trace over the dot,
one arrives at the expression for the current in terms of the new kernel and the reduced density matrix in 
the interaction picture, \Eq{current-ia2}. Transformed back to the Schr\"odinger picture we obtain \Eq{current}.

\subsection{Comparison of the approaches\label{NZ-RTD}}

For the comparison between the \FQME{} and \RT{} approaches, it is most useful to consult 
\Eq{\FQME-explicitkernel} and \Eq{rt-tnl}.
The second order terms, contained in both equations in the first line, obviously match.
The equivalence of the fourth-order terms
is more indirect:
in the \FQME{} approach, the first term of  \Eq{\FQME-explicitkernel} is evaluated using Wicks' theorem by building all possible contractions, including the reducible ones
(contraction between vertices at times $t$ and $\tau_2$ as well as $\tau_1$ and $\tau$).
The latter are then canceled by the second term.
Precisely the same happens in the projection operator approach [cf. \Eq{eq:KIproj4}].
The above conclusions hold in fact for any order of perturbation theory as shown in~\cite{Timm08}.
In contrast, the \RT{} approach avoids the inclusion and subsequent cancellation of reducible 
parts which rapidly grow in number with the perturbation order.
\par
We emphasize that there is one unique correct fourth order (time non-local) generalized master equation,
in which the kernel includes \emph{all fourth order contributions}, but \emph{no higher order ones}
 and which does \emph{not diverge} in the stationary (zero-frequency) limit.
This master equation can be derived using either the \FQME{} or \RT{} (or NZ)
approaches and there is no need to distinguish between these in the following 
discussion.\par
After this comparison on a formal level, we will continue in \Sct{diagr-repr} with a comparison on the level of the individual contributions to the time-evolution kernel.

\section{Diagrammatic representation and mapping between \FQME{} and \RT{}\label{diagr-repr}}
We now address the task of calculating all elements ${{K}}^{aa'}_{bb'}(i0)$ of the \Kkernel{} in the stationary  \GME{}, \Eq{vectoreq-1}.
For our purposes, it will turn out to be advantageous to first work in the time-domain, i.e., to calculate ${\mathcal{K}}^{aa'}_{bb'}(t-\tau)$,
which we decompose into contributions of successive non-vanishing even orders $n=2,4,...$ in the tunnel coupling:
\[(\mathcal{{K}})^{aa'}_{bb'}=(\mathcal{{K}}^{(2)})^{aa'}_{bb'}+(\mathcal{{K}}^{(4)})^{aa'}_{bb'}+\cdots.\]
The section has a twofold aim.
We first introduce the diagrammatic representation for the \Kkernel{} and show how each \FQME{} contribution,
obtained from \Eq{\FQME-explicitkernel}, translates into a corresponding diagram in the \RT{} approach based on \Eq{rt-tnl}.
Apart from being of technical interest, this is of key importance for the discussion in \Sct{nonsec}, where the fourth order kernel elements
incorporating corrections to the diagonal elements of the density matrix due to the non-diagonal elements are introduced.
Secondly, we discuss the time-dependent part of the kernel contributions in the Schr\"odinger picture
and its zero-frequency Laplace transform, on which the simplified calculation of the effective fourth order kernel in \Sct{group} relies.
\par
\begin{figure}
\begin{center}
\begin{tabular}{rl}
\raisebox{0.6cm}{\Large{$\left(\mathcal{K}^{(4)}\right)^{aa'}_{bb'}(t-\tau)=\ $}}&
\includegraphics[width=3cm]{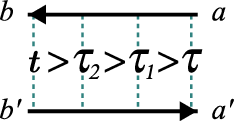}
\end{tabular}\end{center}
\caption{Time ordering in a diagram associated with a fourth order process. Every term arising from the \FQME{} approach, \Eq{\FQME-explicitkernel}, can uniquely be translated into a specific diagram. This gives a one-to-one mapping between the \FQME{} and \RT{} approach. While the time-order is crucial for this mapping, the resulting diagram can also directly be used to represent the Laplace transformed expression (see \Fig{diagr-table}).}\label{4QME-diagr1}
\end{figure}

In the conventional RT approach one starts by considering super matrix elements [cf.~\Eq{supermatel}] of the kernel~\footnote{See, however, the recently introduced superoperator formulation of the real-time approach~\cite{Schoeller09b,Leijnse08a} which does not refer to matrix elements.}
and one introduces a diagrammatic representation for the order $n=2,4,\ldots$ parts of the kernel:
\begin{center}
\begin{tabular}{rl}
\raisebox{0.54cm}{\large{$\bigl(\mathcal{K}^{(n)}\bigr)^{aa'}_{bb'}(t-\tau)=\ $}}& 
\includegraphics[height=1.6cm]{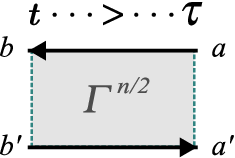}
\end{tabular}
\end{center}
The diagram represents operators which act from the left and right on the dot density operator $\hrho(\tau)$,
inducing an irreducible time-evolution of a pair of \textit{initial} states $a,a'$ [associated with $\hrho(\tau)$] to corresponding \textit{final} states $b,b'$ [belonging to $\dot{\hrho}(t)$].
Time thus increases from right to left.
The diagram can be considered as a Keldysh \textit{contour}, i.e., running from  $a \to b$, then continuing backward from $b' \to a'$, as indicated by the directed line on the upper and lower part.
On the upper (lower) part of the contour the time-ordering  agrees with (is opposite to) the contour direction, indicating that operators acting from the right on the density operator $\hrho(\tau)$ come in inverted order concerning time.
This distinction is important for the diagram rules.
The shaded area indicates the sum of all contributions involving the product of $n$ tunnel operators $H_T$, starting at time $\tau$ and ending at time $t$ (this thus yields a product of $\frac{n}{2}$ broadening elements, which we indicate with $\Gamma^{n/2}$).
Starting from the \RT{} expression for the \Kkernel{}, \Eq{rt-tnl}, simple rules are derived, given in \App{app:diagr_rules}, 
from which one can directly read off the analytical expression for the zero frequency Laplace transform of each diagram.~\cite{Schoeller97hab}\\
Hence, the diagram contains not only the information about the contribution to the kernel $\mathcal{K}$ in the interaction picture, but also to its Laplace transform $K$. To make a distinction, we will use the convention that we mean contribution to $\mathcal{K}$ whenever we explicitly write down time labels in the diagram. Otherwise it stands for the Laplace transformed expression, i.e. the contribution to $K$ (this is the case everywhere in the following except for \Fig{4QME-diagr1}).

In the BR approach one has to expand the superoperator expression~\Eq{\FQME-explicitkernel} in commutators with dot and electrode operators.
One then applies Wick theorem to integrate out the electrodes, resulting in cancellations of terms.
Finally super-matrix elements are taken and the remaining expressions correspond term by term to the \RT{} diagrams.
To emphasize the close connection between the two approaches we now illustrate this explicitly
 by calculating second and fourth order contributions to the \Kkernel{} in the \FQME{} approach.
To this end, we first split the tunneling \noEq{Hamiltonian}{lt_ham} into two parts,
\[ \hbar^{-1} \hat{H}^I_T(\tau_k)=\hat{A}^+_k+\hat{A}^-_k,\]
describing tunneling into ($p=+$) and out of ($p=-$) the dot:
\begin{subequations}\begin{align}
&\hat{A}^{+}_k=:\hat{D}^{+}_k\hat{C}^{-}_k:=\Bigl(\hbar^{-1}\hat{d}^\dag_{\sigma m}(\tau_k)\Bigr)\Bigl(t_{lmq}\,\hat{c}_{l\sigl q}(\tau_k)\Bigr),\label{substitution}\\
&\hat{A}^{-}_k=:\hat{C}^{+}_k\hat{D}^{-}_k:=\Bigl(t^*_{lmq}\,\hat{c}^\dag_{l\sigl q}(\tau_k)\Bigr)\Bigl(\hbar^{-1}\hat{d}_{\sigma m}(\tau_k)\Bigr).\label{substitution2}
\end{align}\end{subequations}
Here the index $k$ numbers the time argument, with $t$ and $\tau$ corresponding to $k=3$ and $k=0$, respectively.
The summations over $l,m,\sigma,q$ are implicitly understood.
We insert this into \Eq{\FQME-explicitkernel}, denoting $\hrho^I_0:=\hrho^I(\tauz)$:
\begin{subequations}\begin{align}
&\hat{\mathcal{K}}^{(2)}(t-\tauz)\hrho^I(\tauz)=-\sum_{\substack{p_0,p_3\\\in\{+,-\}}}\Tr_\leads\Bigl[\,\hat{A}^{p_3}_3,\bigl[\,\hat{A}^{p_0}_0,\hrho^I_{0}\hrho_\leads\,\bigr]\,\Bigr],\label{dbl-comm}
\\
&\hat{\mathcal{K}}^{(4)}(t-\tauz)\hrho^I(\tauz)=\sum_{p_0,p_1,p_2,p_3\in\{+,-\}}\,\timeint{t}{\tau_2}{\tau_1}{\tau}\times\nonumber\\&\Bigl(\ \Tr_\leads\Bigl[\,\hat{A}^{p_3}_3,\Bigl[\,\hat{A}^{p_2}_2,\bigl[\,\hat{A}^{p_1}_1,\bigl[\,\hat{A}^{p_0}_0,\hrho^I_{0}\hrho_\leads\,\bigr]\,\bigr]\,\Bigr]\,\Bigr]\Bigr.\nonumber\\\Bigl.-&\Tr_\leads\Bigl[\,\hat{A}^{p_3}_3,\Bigl[\,\hat{A}^{p_2}_2,\Tr_\leads\left\{\bigl[\,\hat{A}^{p_1}_1,\bigl[\,\hat{A}^{p_0}_0,\hrho^I_{0}\hrho_\leads\,\bigr]\,\bigr]\right\}\hrho_\leads\,\Bigr]\,\Bigr]\ \Bigr).\label{quatr-comm}
\end{align}\end{subequations}
We next expand the  multiple commutators and
collect the fermionic operators of the leads, using that they anti-commute with the quantum dot operators,
 $\hat{C}^{p_k}_k\hat{D}^{\bar{p}_k}_k=-\hat{D}^{\bar{p}_k}_k\hat{C}^{p_k}_k$, where $\bar{p}_k = -p_k$.
\begin{figure*}
\includegraphics[width=\textwidth]{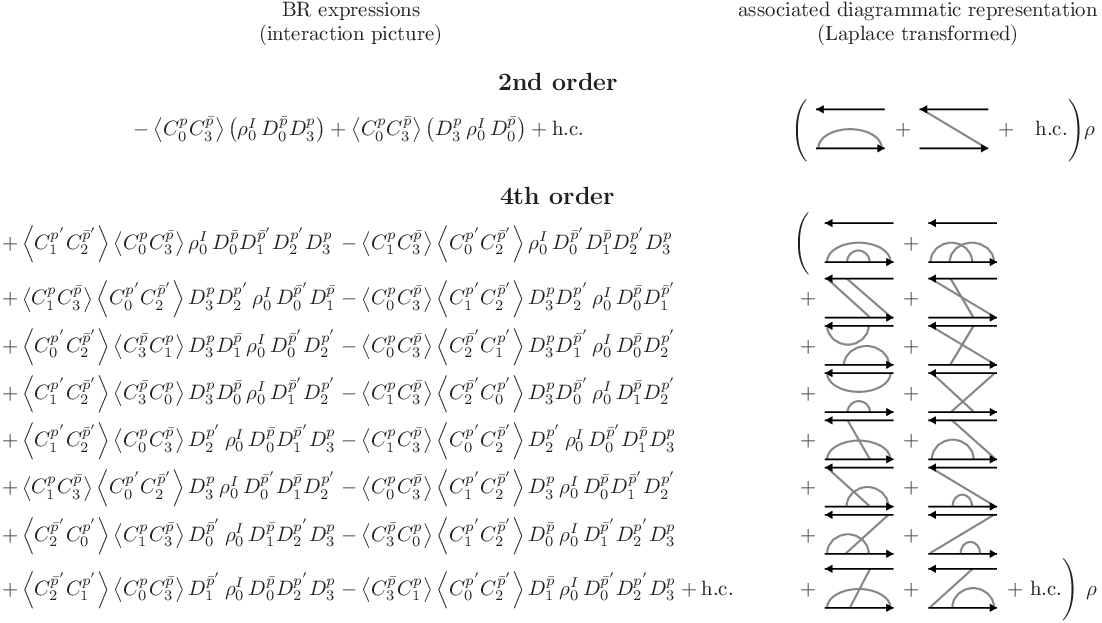}\caption{On the left we give the contributions to the \Kkernel{} $\mathcal{K}^{(2)}$ and $\mathcal{K}^{(4)}$ as they arise from the \FQME{} in the interaction picture. Here, $p=\pm,\,\bar{p}=-p$. The corresponding diagrammatic \RT{} representations, standing directly for the Laplace transformed contributions to $K^{(2)}$ and $K^{(4)}$, are shown on the right. All minus signs of the \FQME{} contributions are incorporated into the diagrams.}\label{diagr-table}\end{figure*}
Using the cyclic property of the trace and Wick's theorem, the average of the lead fermionic operators
is expressed as  a sum over products of pair-contractions. This results for \Eq{dbl-comm} and \Eq{quatr-comm} in the expressions listed on the left sides of \Fig{diagr-table}. For the fourth order, the reducible contractions emerging from the last two terms in \Eq{\FQME-explicitkernel} cancel each other.
Using the time-ordering of \Fig{4QME-diagr1}, \Fig{diagr-table} further gives on the right sides the respective \RT{} diagram for each expression.
The Hermitian conjugated terms, which have been omitted in the figures, correspond to diagrams which are vertically mirrored, i.e. all vertices on the upper \contourorpropagator{} have to be moved to the lower one and vice versa.

The translation from the \FQME{} expressions in the interaction picture into the \RT{} diagrams works as follows.
For each operator $D^{p_k}_k$ standing on the left (right) of $\hrho^I_0$, draw a vertex on the upper (lower) contour at time $\tau_k$.
For each contraction $\langle C^{\bar{p}_i}_i C^{\bar{p}_j}_j \rangle$, requiring $p_i=-p_j$ in order not to vanish,
draw a contraction line between the vertices representing $D^{p_i}_i$ and $D^{p_j}_j$.
Notice that the ordering of the $C$-operators in each contraction is consistently incorporated in the diagram:
the pair of $C$ operators in the contraction have the same time-ordering as the corresponding $D$ operators, unless the earlier vertex of the two lies on the lower part of the contour (this follows from the cyclic permutation under the trace).
Similarly, the sign of the operator expression, 
$(-1)^{n/2+ n_\text{c} + n_\text{l}}$,
is automatically contained in the diagram through the number of contractions $n/2$ ($n$=order of perturbation theory), the number of crossing contraction lines $n_\text{c}$, and the number of vertices on the lower contour $n_\text{l}$.

\begin{figure}
\includegraphics[width=0.98\columnwidth]{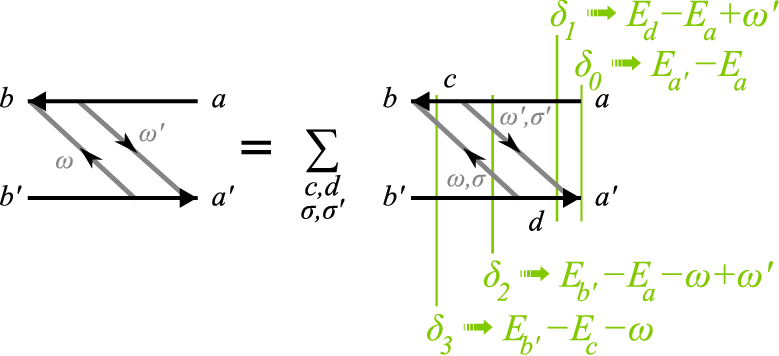}
\caption{Example of one possible fourth order contribution to the kernel element $K^{a a'}_{b b'}$. Here, $\omega$ and $\omega'$ are the energies which are assigned to the fermion lines. Unless the intermediate states of a diagram are labelled (like for example in \Sct{nonsec} concerning reducible diagrams), it is implicitly meant to contain the sum over any intermediate state. As described in the main text, the quantities $\delta_i$ can by read off from cuts through the diagram. For our example: $\delta_0=E_{a'}-E_a$, $\delta_1=\rmi 0+E_d-E_a+\omega'$, $\delta_2=\rmi 0+E_{b'}-E_a-\omega+\omega'$, $\delta_2=\rmi 0+E_{b'}-E_c-\omega$.
\label{deltacuts}}
\end{figure}

The diagrams listed in \Fig{diagr-table} represent expressions which are summed over the indices $p_k=\pm$.
Terms with specific values of the $p$'s are represented by diagrams where the contraction lines are directed by an arrow,
pointing towards the vertex corresponding to $D^{+}$.
Figure~\ref{deltacuts} shows an example for the third 4th order diagram in \Fig{diagr-table}.
From the diagrams it is explicitly clear that all contributions which where not canceled are irreducible:
between the first and the last vertex at times $t$, respectively $\tau$, there is no time point at which the diagram could be separated into two parts without cutting a contraction line.
\par
To obtain the stationary kernel \Eq{\FQME-explicitkernel} required in \Eq{rho-lapl} we first need to transform back to the Schr\"odinger picture [cf.~\Eq{time-lop}]
by inserting\begin{subequations}
\begin{eqnarray}
\hat{D}^{\pm}_k&\sim&\hat{d}^{(\dag)}_{\sigma m}(\tau_k)=e^{\frac{\rmi}{\hbar} \hat{H}\tau_k}\hat{d}^{(\dag)}_{\sigma m}e^{-\frac{\rmi}{\hbar}\hat{H}\tau_k},\label{transD}\\
\hat{C}^{\pm}_k&\sim&\hat{c}^{(\dag)}_{l\sigma q}(\tau_k)=e^{\frac{\rmi}{\hbar} \hat{H}_\leads\tau_k}\hat{c}^{(\dag)}_{l\sigma q}e^{-\frac{\rmi}{\hbar} \hat{H}_\leads\tau_k},\label{transC}\\
&&\hrho^I(\tau_k)=e^{\frac{\rmi}{\hbar} \hat{H}\tau_k}\hrho(\tau_k) e^{-\frac{\rmi}{\hbar} \hat{H}\tau_k}.\label{transrho}
\end{eqnarray}\end{subequations}
For the further developments in this paper only the time-dependent part of the resulting expression, and its Laplace transform, is of importance.
We can only factor out this part after taking super-matrix elements [cf.~\Eq{supermatel}] of the kernel contributions with respect to the energy-eigenstates of the quantum dot
and insert complete sets of these states between all the vertex operators $D$.
The resulting expression is represented by a diagram labeled with these dot eigenstates on the contour, as illustrated in \Fig{deltacuts}.
We calculate its Laplace transform with respect to $\tau'=t-\tau$ ($ \tau_{n-1}=t, \tau_{0}=\tau$),
collecting for each time $\tau_k$ all energy contributions into one exponential with argument $-i\Delta_k\tau_k/\hbar$:
\begin{align}
\left( K(z) \right)^{a a'}_{b b'} \sim
\int_0^\infty\!\!\rmd\tau'e^{\frac{\rmi}{\hbar}\Delta_0\tau'}
\timeint{t}{\tau_{n-2}\cdots}{\tau_1}{t-\tau'}\prod_{k=1}^{n-1}e^{-\frac{\rmi}{\hbar}\Delta_k\tau_k}
\end{align}
Here the additional $\Delta_0$ contains both the Laplace variable $z=i0$ and the energy difference
$E_{a'}-E_a$ of the initial states on the upper and lower part of the contour.
Transforming variables to the time-differences between vertices $\tilde{\tau}_k:=\tau_{k-1}-\tau_{k}$
decouples the integrals,
showing that the energies $\delta_{k}=\sum_{l=0}^k\Delta_l$ fully determine the time-evolution factor and its zero-frequency transform:
\begin{align}
\label{expr1}
\left( K(z) \right)^{a a'}_{b b'} \sim \prod_{k=1}^{n-1}\int_0^\infty\!\!\rmd\tilde{\tau}_k\, e^{-\frac{\rmi}{\hbar}\delta_{k}\tilde{\tau}_k}
=\prod_{k=1}^{n-1}\frac{1}{\delta_k},
\end{align}
This is the form of the zero frequency Laplace transform of the time-dependent factor only, 
as obtained from the diagram rules in \App{app:diagr_rules}, which is the most convenient starting point for our analysis.
The energy $\delta_k$ is obtained from a diagram by making a vertical cut through the diagram between times $\tau_k$ and $\tau_{k-1}$,
and then adding to / subtracting from $z$
the energies of all directed lines which hit this cut  from the left / right.
This includes the energies associated with contraction lines as well as the upper and lower part of the contour.
Figure~\ref{deltacuts} demonstrates how this simple rule works for a specific fourth order diagram.
A crucial point for the rest of the paper is that diagrams which differ by breaking time-ordering \emph{between the contours}, but keeping  
time-ordering \emph{within each contour}, only differ by the arguments of the time-dependent exponentials in \Eq{expr1}, i.e., in the $\delta_k$.
As \App{app:diagr_rules} shows, the products of TMEs and  electrode distribution functions and the overall phase factor are identical.
The simplifications discussed below are thus independent of these factors and their precise form needs no further discussion.

\section{Coherences and non-secular corrections\label{nonsec}}
For many simple quantum dot models, selection rules deriving from symmetries prevent the occupations
of the dot states from coupling to the coherences. 
Whenever two states $a$ and $a'$ of the system differ by some quantum number which is conserved in the
\emph{total} system (i.e. including the reservoirs),
then their coherence $\rho_{aa'}$  does not enter into the calculation of the occupancies since
$K^{a a'}_{bb}=0$ for all states.
The simplest example for such a quantum number is the electron number which is conserved for a quantum dot coupled to  non-superconducting electrodes.
The total spin-projection is also conserved for unpolarized or collinearly polarized electrodes.
For non-collinear polarizations, however, inclusion of the coherences is crucial in order to capture spin-precession effects~\cite{Braun04set, Braun06rev}.
In a similar way orbital degeneracies have been shown to affect the occupations through the 
coherences.~\cite{Begemann08,Donarini09} So in general, coherences cannot be neglected. Making no specific assumptions about the coherences, the only selection rule we enforce here is the one due to 
the conservation of total charge.
In the above mentioned works, the tunneling is treated to lowest order $(\hat{\mathcal{L}}_T^2)$
and only non-diagonal elements between degenerate states are kept.
The latter so-called \textit{secular approximation} is usually phrased as
neglecting the rapidly oscillating terms corresponding to
coherences between non-degenerate states.~\cite{Blum_book}

In fourth order, however, an elimination of coherences between non-degenerate states in the stationary limit requires an expansion of the effective kernel for the occupations.
Such an expansion is consistent in the sense that the derived effective kernel includes all contributions up to fourth order (while a simple neglect of non-secular coherences introduces serious errors of the order $\hat{\mathcal{L}}_T^4$).~\cite{Leijnse08a,koller_thesis}

We start by decomposing the density matrix into 
a \emph{secular} (energy diagonal) part $\hrho_{s}$ and a
non-secular (energy non-diagonal) part $\hrho_{n}$.
Here $\hrho_{n}$ contains all matrix elements $\rho_{aa'}$ between states with 
$\left| E_a - E_{a'} \right| > \epsilon_n$ and $\hrho_{s}$ all other elements (including the diagonal components, $a = a'$, 
corresponding to the populations).
The cutoff $\epsilon_n$ should be chosen large compared to the tunnel broadening of the states, $\epsilon_n \gg \hbar\Gamma$,
the precise requirement being that it should be large enough that the next-to-leading order term in the expansion of \Eq{eom-nonsec} below 
is comparable to a sixth order term, and can thus be neglected.
Our aim is to eliminate the non-secular coherences $\hrho_{n}$ and include their effect as a correction to the kernel determining the 
secular part.
To this end we write \Eq{rho-lapl} in block matrix form,
\[
\left(
\!\begin{array}{c}0\\0\end{array}
\!\right)
\!=\!
\left(\!\!\!
\begin{array}{c@{\,}c} -\rmi \mathcal{L}_{ss}\!+\!\hat{{K}}^{(2)}_{ss}\!+\!\hat{{K}}^{(4)}_{ss}&\hat{{K}}^{(2)}_{sn}\!+\!\hat{{K}}^{(4)}_{sn}\\\hat{{K}}^{(2)}_{ns}\!+\!\hat{{K}}^{(4)}_{ns}& -\rmi \hat{\mathcal{L}}_{nn} \!+\!\hat{{K}}^{(2)}_{nn}\!+\!\hat{{K}}^{(4)}_{nn}\end{array}\!\!\!\right)\!\left(\!\!\begin{array}{c}\hrho_{s}
\\
\hrho_{n}\end{array}\!\!\right),
\]
where, see \Eq{vectoreq-1}, the free evolution of the system involving $(\hat{\mathcal{L}})^{aa'}_{bb'} \equiv (E_{a'}-E_a) \delta_{ab} \delta_{a'b'}$ is zero in the $ns$ and $sn$ blocks by definition.
Solving for $\hrho_{n}$ one obtains
\begin{equation}
\hrho_{n}=-\left( - \rmi \hat{\mathcal{L}}_{nn}+
\hat{{K}}_{nn}^{(2)}+\hat{{K}}_{nn}^{(4)}\right)^{-1}\!\left(\hat{{K}}^{(2)}_{ns}+\hat{{K}}^{(4)}_{ns}\right)\hrho_{s},\label{eom-nonsec}
\end{equation}
which obviously contains all orders in $\Gamma$ due to the inverse. Since by construction $ \mathcal{L}_{nn} \gg \hbar\Gamma $
we can expand \Eq{eom-nonsec} in $\hbar\Gamma / \mathcal{L}_{nn}$, finding that the lowest order term gives corrections to the secular density matrix of order 
$\Gamma^2$ and is thus all that should be kept in a consistent fourth order expansion.
Inserting this in the equation for the secular part of the density matrix
we obtain an effective stationary kinetic equation for the secular density matrix:
\begin{equation}
  \label{eff}
  0=\left(- \rmi \hat{\mathcal{L}}_{ss}+\hat{{K}}^{(2)}_{ss}+\hat{{K}}^{(4)}_{\eff} \right)\hrho_{s},
\end{equation}
with the effective fourth order kernel
\begin{align}
  \hat{{K}}^{(4)}_\eff \equiv \hat{K}_{ss}^{(4)}+\hat{K}^{(4)}_\text{N}.
  \label{Keff}
\end{align}
Here\begin{equation}
  \label{diagramcorrections}
  \hat{{K}}^{(4)}_\text{N} = \hat{{K}}^{(2)}_{sn}\frac{\rmi}{-\hat{\mathcal{L}}_{nn}}  \hat{{K}}^{(2)}_{ns}
\end{equation}
is the correction to the secular density matrix due to coherences between non-secular states.
This makes explicit that when going beyond lowest order, the secular approximation is no longer valid and also coherences 
between non-secular states have to be accounted for.
This was shown in~\cite{Leijnse08a} for 
the special case where the secular part is diagonal.
Here we extended the derivation to an arbitrary excitation spectrum, where the secular part may be non-diagonal,
i.e., the effective equation is \emph{not} a master-equation for occupancies.
It should be noted that in this case the kernel ${K}^{(4)}_{ss}$ must be calculated including the elements which couple to secular coherences.

For the developments of the present paper it is useful to introduce a diagrammatic representation of the non-secular correction $\hat{{K}}^{(4)}_\text{N}$.
We first note that the inverse of $\hat{\mathcal{L}}_{nn}$ is related to a diagram without tunneling lines by the diagram rules, see \App{app:diagr_rules}, evaluated at zero frequency ($z=\rmi 0$):
\begin{center}\vskip -0.2cm\begin{tabular}{l@{\hspace{-0.3cm}}r}
\begin{minipage}{6cm}
\[\left( \frac{\rmi}{-\hat{\mathcal{L}}_{nn}} \right)^{aa'}_{aa'}\ =\ \frac{\rmi}{E_{a'}-E_{a}}\ 
\equiv\]\vspace{1mm}\ \end{minipage}&\begin{minipage}{4cm}
\includegraphics[height=1.2cm]{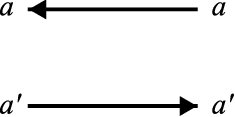}
\end{minipage}\end{tabular}\vskip -0.2cm\end{center}
Note that this "free evolution" term is always finite since the expansion is only carried out in the 
non-secular subspace where $|E_a-E_{a'}| \gg \hbar\Gamma$ and $\rmi 0$ can always be dropped.
Representing a general second order contribution diagrammatically as
\begin{center}\vskip -0.2cm\begin{tabular}{l@{\hspace{-0.3cm}}r}
\begin{minipage}{3.4cm}
\[\left(\hat{{K}}^{(2)}\right)^{aa'}_{bb'}\ \equiv\]
\vspace{0.3mm}\
\end{minipage}&\begin{minipage}{3.4cm}
\includegraphics[height=1.2cm]{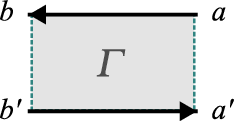}
\end{minipage}\end{tabular}\vskip -0.2cm\end{center}
the correction term due to coherences between non-secular states is given by 
\begin{center}\vskip -0.2cm\begin{tabular}{l@{\hspace{-0.2cm}}r}
\begin{minipage}{4.2cm}
\[\left(\hat{{K}}^{(4)}_\text{N}\right)^{aa'}_{bb'}\ = \sum_{c,d \in n} \]\vspace{1mm}\
\end{minipage}&
\begin{minipage}{2cm}
\includegraphics[height=1.6cm]{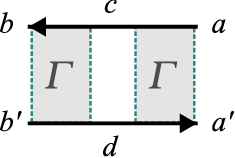}
\end{minipage}\end{tabular}\vskip -0.2cm\end{center} 
where the sum is restricted to states which are pairwise non-secular, i.e., $c$ and $d$ with $|E_d - E_c| \gg \hbar\Gamma$.

Thus, $\hat{{K}}^{(4)}_\text{N}$ appears as a sum of all \textit{reducible} fourth order diagrams with non-secular intermediate free propagating states $c,d$. 
The evaluation can be performed by using the diagram rules, \App{app:diagr_rules}, as for an \textit{irreducible} fourth order diagram.
The effective fourth-order part of the kernel, determining the secular part of the stationary density matrix through \Eq{eff},
can thus be calculated in the same way as $K^{(4)}$, with only the following  modifications of the diagram rules:
(i) diagrams can be irreducible and reducible \emph{between the first and last vertex};
(ii) the intermediate states of reducible diagrams are restricted to non-secular free propagating intermediate states;
(iii) only secular initial and secular final states for the diagrams are possible.
We finally note that when calculating the current using the current kernel,
the contributions of the non-secular coherences can be eliminated in exactly the same way.
The only modification required is to replace the operator acting at the latest time by the corresponding current operator.

We have thus eliminated the non-secular coherences from the transport calculation.
This effective diagrammatic theory for the secular part of the density matrix is the starting point for 
the evaluation of diagrams in \textit{groups}, rather than single ones, which we turn to now.
This will result in simplifications of the numerical evaluation of the kernels (\Sct{group}),
and allow the relation to the \Tmat{} approach to be established (\Sct{T-mat}).

\section{Efficient diagram evaluation\label{group}}
\begin{figure}
\begin{center}
\includegraphics[width=0.76\columnwidth]{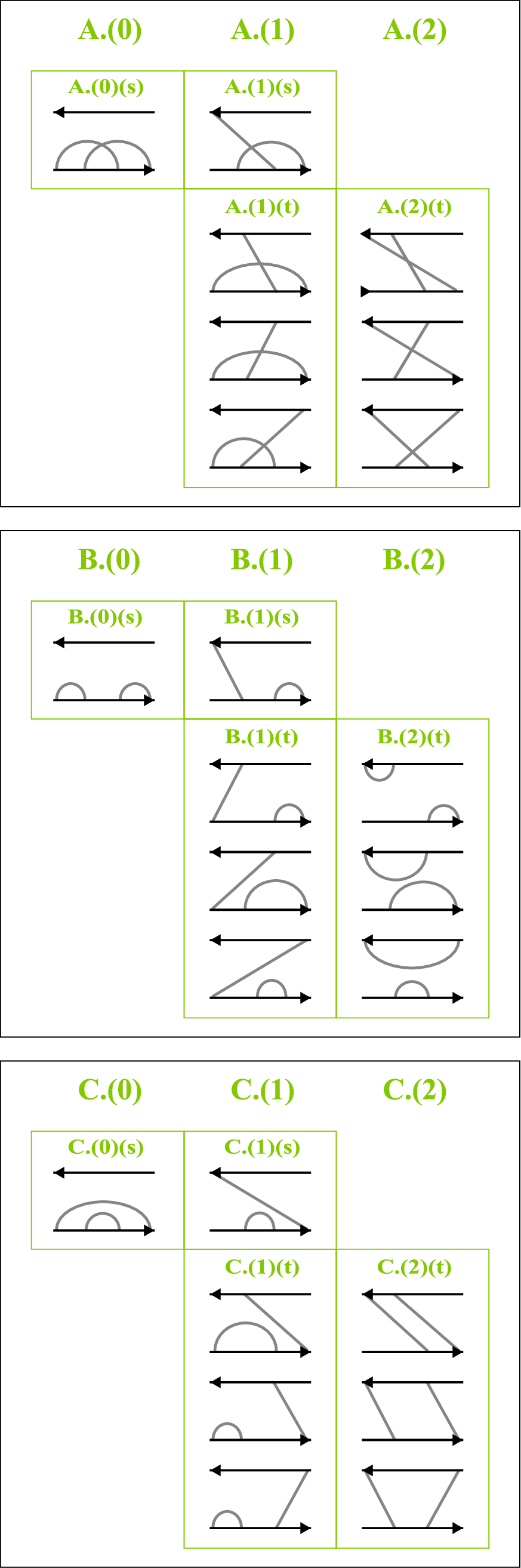}
\end{center}
\caption{The sixteen irreducible fourth order diagrams, together with the eight reducible correction diagrams, 
can be sorted by topology into three diagram classes G = A, B, C. Within each class, there are 
three groups \sg.(0), \sg.(1), \sg.(2), labeled by the number of vertices on the upper part of the contour.
In \sg.(1), the latest (leftmost) vertex distinguishes between stand-alone diagrams (s) and triple diagram subgroups (t).}\label{diagr-tmat}
\end{figure}
We now focus on the explicit evaluation of the effective fourth order transport kernel \eq{eff} in the zero-frequency limit,
\begin{align}
  \hat{{K}}^{(4)}_\eff\equiv {K}_{ss}^{(4)}+\hat{K}^{(4)}_\text{N},
  \label{eff2}
\end{align}
according to the  modified diagram rules derived above.
For each super matrix element of this kernel [cf.~\Eq{supermatel}] 
this involves evaluation of all fourth order diagrams
for all possible combinations of all intermediate states on both propagators.
Already for quantum dot models with a moderate number of states $\sim 10-100$ this comes at a high numerical cost.
In this section we demonstrate a way to reduce the computational cost drastically \emph{without introducing any approximation}.
\par
In \Fig{diagr-tmat} we show the 24 diagrams representing the 16 irreducible and 8 reducible diagrams of $K_{ss}^{(4)}$ and $K_\text{N}^{(4)}$, respectively.
These represent all 192 contributions to the effective kernel \Eq{eff2},
since we do not specify the direction of the two contraction lines ($p$ indices)
and only include one diagram from each hermitian conjugate pair.
It is only by combining diagrams from both $K_{ss}^{(4)}$ and $K_\text{N}^{(4)}$, which is necessary as explained in \Sct{nonsec},
that a structure is revealed upon which our efficient diagram evaluation is based.
The diagrams are sorted in~\Fig{diagr-tmat} in three steps.

First, there are three diagram classes, \sg$\in\{$A, B, C$\}$.
These distinguish the three topologically different ways in which the four vertices can be contracted,
considering them to lie on the \emph{contour} (i.e. moving vertices between the upper and lower part of the contour via the latest time does not alter the topology.)

Secondly, these classes divide into groups \sg.($x$),\ $x\in\{0,1,2\}$, based on the number of vertices $x$ on the upper
part of the contour.
(Diagrams with $x=\{3, 4\}$ need not be included in \Fig{diagr-tmat} since they are hermitian conjugates to diagrams in the $x=\{1, 0\}$ groups.)
The classes are thus constructed by forming the one-member group \sg.($0$) where all vertices lie on one part of contour (here taken to be the lower part)
 and then successively shifting the vertices to the opposite part of the contour via the latest time point.

Thirdly, one distinguishes subgroups by the position of the latest vertex, being either on the upper or lower part of the contour:
the groups \sg.$(1)$ thus divide into a subgroup \sg$.(1)$(s) (single) and \sg.$(1)$(t) (triple) of one and three diagrams respectively,
whereas in the groups \sg.$(0)$ and \sg.$(2)$ there exists only the one-diagram respectively three-diagram subgroup, such that \sg.$(0)\equiv$\sg.$(0)$(s) and \sg.$(2)\equiv$\sg.$(2)$(t).
\par
A key point is that diagrams within a group \sg.($x$) give rise to expressions which in the interaction picture only differ by their time-dependent factor, and hence in Laplace space only differ by their frequency dependent part.
They transform into one another by \emph{breaking time-ordering between the different parts of the contour}, i.e.,
freely shifting around vertices  without breaking time-ordering \emph{on each part of the contour}.
It follows from the diagram rules that they all come with the same TMEs, electrode distribution functions, and overall sign, cf.~\Sct{diagr-repr}.
By considering only the time-dependent part and its Laplace transform, we derive in~\Sct{triple} new diagram rules for evaluating an entire subgroup at once, arriving at an expression as simple as that for a single diagram.
This halves the number of diagrams one needs to evaluate -- and actually the number can even be halved once more. How to achieve this is explained in~\Sct{gain-loss}: 
One can exploit that each diagram within a class is related to its horizontal neighbor in~\Fig{diagr-tmat} by moving the latest vertex up or down.
Together with a diagram-based (rather than rate-based) looping this results in a speedup by a factor of 10-20 in actual numerical calculations.
Finally, in~\Sct{physics} we explain in what way the classes contribute to effective rates for different physical processes
and illustrate the importance of this in the various transport regimes defined by the applied bias and gate voltage.

\subsection{Evaluating subgroups of diagrams\label{triple}}
We start the efficient evaluation of the sum of diagrams of a subgroup, \sg.($x$)(t),
by selecting a representative diagram and labeling the times $\tau_k$ in all diagrams in the subgroup \emph{based on this diagram}.
Here we take the topmost diagram in \Fig{diagr-tmat} in each subgroup  \sg.($x$)(t), where all the vertices on the upper part of the contour are positioned at the latest possible times (as far as the subgroup allows for this).
This choice is advantageous for the further developments in \Sct{gain-loss}.
We read off the energy differences $\delta_k$ from this representative diagram only, and Laplace transform the time-dependent part of this individual fourth order diagram [cf. \Eq{expr1}]:
\begin{multline} \label{eq:one_diagram}
\timeintthree{t}{\tau_{2}}{\tau_{1}}{\tau}{-\infty}
\prod_{k=0}^{2}e^{\frac{\rmi}{\hbar}(\tau_{k+1}-\tau_k)\delta_{k+1}}\\=
\prod_{k=1}^{3} \int_0^\infty\rmd\tilde{\tau}_k e^{-\frac{\rmi}{\hbar}\delta_{k}\tilde{\tau}_k} \sim
\prod_{k=1}^{3}\frac{1}{\delta_k},
\end{multline}
The other diagrams within the subgroup \sg.($x$)(t) are related by breaking the time-ordering between vertices  on different parts of the contour,
but keeping the position of the latest vertex fixed at time $t$.
For our choice of the representative diagram, this is equivalent to letting the vertex with time $\tau_2$ move freely.
Summing over the three diagrams then exactly corresponds to freely integrating over $\tau_2$, as is shown in \App{app:grouping}.
Thus the zero-frequency Laplace transform of the \textit{sum of all diagrams} within a subgroup \sg.$(x)$(t) is given by
\begin{subequations}
\begin{multline}\label{expr1_triple}\mathrm{G}.(x)(\mathrm{t})\sim\int_{-\infty}^t\rmd\tau_2\timeint{t}{\tau_1}{\tau}{t_0}
\prod_{k=0}^{2}e^{\frac{\rmi}{\hbar}(\tau_{k+1}-\tau_k)\delta_{k+1}}\\
=\int_0^\infty\rmd\tilde{\tau}_3 e^{-\frac{\rmi}{\hbar}(\delta_3-\delta_2)\tilde{\tau}_3}\prod_{k=1}^{2}
\int_0^\infty\rmd\tilde{\tau}_k e^{-\frac{\rmi}{\hbar}\delta_{k}\tilde{\tau}_k}\\\sim\frac{1}{\delta_3-\delta_2+\rmi0}
\prod_{k=1}^{2}\frac{1}{\delta_k}.
\end{multline}
This result is just as simple as for a single diagram; the only difference 
between~\Eq{expr1_triple} and~\Eq{eq:one_diagram} is the energy appearing in the leftmost 
denominator.
This is a central result of the paper:
we directly obtain the contribution of a whole subgroup by modifying the diagram rule for the zero-frequency propagator.
One has to evaluate only the representative diagram and assign to the latest segment the propagator $(\delta_3 - \delta_2+i0)^{-1}$ (instead of the usual $\delta_3^{-1}$).
This simplification only works under two conditions:
(i) we are in the zero-frequency limit $z \to i0$ and
(ii) all secular states are degenerate in energy: either $E_a-E_{a'} \gg \hbar\Gamma$ (non-secular) or $E_a-E_{a'} \ll \hbar\Gamma$ (secular = degenerate) holds, i.e., $\mathcal{L}_{ss}$ can be set to zero.
In \App{app:grouping} we show in detail how these conditions enter, in particular the proper handling of imaginary convergence factors $i0$,
 and we discuss a worked out example for subgroup C.($x$)(t).
\par
A point which still requires separate care is the secular cases of the reducible diagrams in classes B and C.
When integrating freely over $\tau_2$ we sum over all diagrams, including the reducible ones.
As discussed in \Sct{nonsec} this should only be done when the intermediate states on the free 
propagator part are non-secular, i.e., when $\delta_2 \neq \rmi 0$ in B.(1/2)(t) and when 
$\delta_1 + \delta_3 \neq \delta_2+\rmi 0$ in C.(1/2)(t).
For intermediate dot states for which this condition is not satisfied we must sum up only the irreducible contributions.
However, similarly to the non-secular case, this can be effected by a non time-ordered integration over $\tau_2$.
For B.(1)(t) and B.(2)(t) two irreducible diagrams remain to be summed for the secular case $\delta_2=\rmi 0$:
\begin{equation}\label{expr1_B}\mathrm{B}.(x)(\mathrm{t})\sim\!\!\!\!\timeint{t}{\tau_1}{\tau}{-\infty}\int_{-\infty}^{\tau_1}
\!\!\rmd\tau_2\prod_{k=0}^{2}e^{\frac{\rmi}{\hbar}(\tau_{k+1}-\tau_k)\delta_{k+1}}
\sim \frac{1}{\delta_3^2 \delta_1}.
\end{equation}\\
The modified diagram rule in this case requires $\delta_3^{-1}$ for the center propagator (instead of the usual $\delta_2^{-1}$).
Note that the energies $\delta_k$, $k=1,2,3$, are those of the \emph{reducible} representative diagram, which is actually excluded from the sum.
For groups C.(1)(t) and C.(2)(t) only the irreducible representative diagram remains in the secular case $\delta_2=\delta_1+\delta_3-\rmi 0$, and the standard rule gives
\begin{equation}\label{expr1_C}\mathrm{C}.(x)(\mathrm{t})
\sim
\frac{1}{\delta_3\delta_2 \delta_1}.
\end{equation}
\end{subequations}
The results~\eq{expr1_triple}--\eq{expr1_C} can alternatively be obtained by directly summing the Laplace transformed propagators of the individual diagrams.
This is shown in~\App{app:grouping} using general relations between the energy denominators of diagrams within a subgroup in the zero-frequency limit.
This could be of use in diagrammatic calculations of quantities other than the density matrix and the current, e.g. current noise~\cite{Thielmann05} and time-dependent observables~\cite{Splettstoesser06}.
Finally, we note that for analytic calculations one can further sum up the four contributions from each \emph{group} \sg.($1$) to a single expression as well, which is, however, of no further advantage for the numerical implementation envisaged here, since one exploits the relations \eq{gl_s}, \eq{gl_t} as explained in subsection \ref{gain-loss}.
\par
The central result \Eq{expr1_triple} can be generalized to any order of perturbation theory $n$,
resulting in a relative computational gain which grows with $n$ (see \App{app:grouping}). 
By the same three step procedure as outlined for the fourth order, the diagrams can be combined into subgroups with $x$ vertices on the upper part of the contour ($x=0,\ldots,n/2$) and $y=n-x$ on the lower one.
All diagrams in the subgroup are generated by moving vertices around on the upper and lower part of the contour, while keeping the contractions and the vertex at the latest time $t = \tau_{n - 1}$ fixed.
We sum over all diagrams in the subgroup by breaking the relative time ordering of the vertices on the different parts of the contour:
\begin{multline}
\timeint{t}{\tau_{n-2}\cdots}{\tau_y}{-\infty}\;\timeint{t}{\tau_{y-1}\cdots}{\tau}{-\infty}\prod_{k=0}^{n-2}e^{\frac{\rmi}{\hbar}(\tau_{k+1} - \tau_k)\delta_{k+1}}\\=
\prod_{k=y+1}^{n-1}\int_0^\infty\rmd\tilde{\tau}_k e^{-\frac{\rmi}{\hbar}(\delta_k-\delta_y)\tilde{\tau}_k}\prod_{j=1}^{y}\int_0^\infty\rmd\tilde{\tau}_j e^{-\frac{\rmi}{\hbar}\delta_{j}\tilde{\tau}_j}\\\sim\prod_{k=y+1}^{n-1}\frac{1}{\delta_k-\delta_y+\rmi0}\prod_{j=1}^{y}\frac{1}{\delta_j}.\label{diagr_rule}
\end{multline}
The subgroup can thus be summed by using the following modified diagram rule.
Determine the propagators for the representative diagram as usual.
Subtract the energy difference $\delta_y$ of segment $y$ from all later ones,
$\delta_k \to \delta_k - \delta_y$, $k>y$.
Here $\delta_y$ belongs to the segment separating the part of the representative diagram with vertices only on the upper and lower part of the contour respectively (ignoring the fixed latest vertex).
The systematic exclusion of reducible diagrams with secular intermediate states (cf. discussion of class B and C in fourth order above)
 can be done most easily in Laplace space by extending the method presented in \App{app:grouping}.

\subsection{Gain-loss relations between diagram groups
and diagram-group based looping\label{gain-loss}}
\begin{figure}
\begin{center}
\includegraphics[width=0.86\columnwidth]{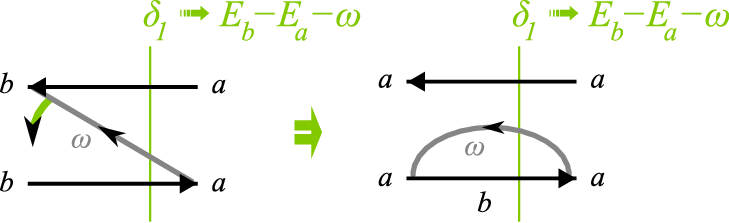}
\caption{Example of the construction of gain and loss partners in second order. From the diagram rules it follows that the contribution of both diagrams are identical \emph{except} for their sign.}\label{gl-2}
\end{center}
\end{figure}
We now shortly address another relation between diagrams, which can be exploited to increase the 
efficiency of their evaluation by another factor two.
Each diagram in \Fig{diagr-tmat} has a horizontal neighbor, 
which is identical except for having the last vertex on the opposite part of the contour.
We will refer to such a pair of diagrams as gain-loss-partners, for reasons which become clear in the following.
Considering a fixed set of intermediate states and assuming the final states to be secular (i.e., either the same or energetically degenerate),
it is easily verified from the diagram rules (see~\Sct{app:diagr_rules}) that
moving the latest vertex to the opposite part of the contour gives the same analytic expression, but with the opposite 
sign.
This is illustrated in  \Fig{gl-2}.
This property of pairs of diagrams implies the sum rule \eq{sumrule} for the kernel~\cite{Schoeller97hab} (which guarantees probability conservation of the density matrix), but is not equivalent to it.
In second order for diagonal diagrams as in \Fig{gl-2} the property has a simple
intuitive interpretation: a tunneling event which changes the dot state from a state $a$ to a state $b$
\emph{increases} the rate by which the occupation probability of the final state $b$ changes, while it \emph{decreases} the rate of change of occupation probability of the initial state $a$.
 The rate for gaining probability in state $b$ is described by the left diagram in
\Fig{gl-2}, adding to the kernel element $K^{aa}_{bb}$. Its partner, the right diagram in \Fig{gl-2}, is obtained by moving the latest
vertex to the opposite part of the contour and
gives the related rate of probability ``loss'' for state $a$. Notice that it adds to a
different kernel element (namely $K^{aa}_{aa}$) exactly the negative of the ``gain'' contribution.
For numerical calculations this implies that if one simply loops over all possible combinations of initial and final states,
the same quantity is calculated twice as a contribution to two different kernel elements.

\begin{center}
\begin{figure}
\includegraphics[width=0.9\columnwidth]{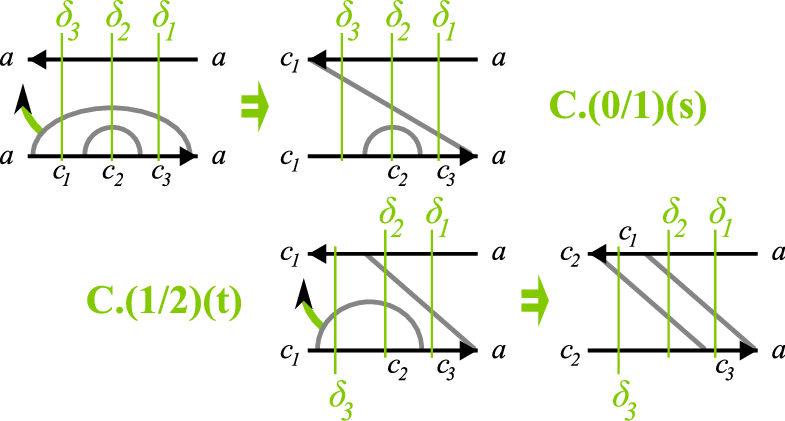}
\caption{Example of the gain-loss-chain among class C diagrams in fourth order.}\label{gl-4}
\end{figure}
\end{center}

This can be avoided: for problems where only diagonal kernel elements $(\hat{K}^{(4)}_\eff)^{aa}_{bb}$
need to be determined, one has merely to calculate \sg.(0)(s) and \sg.(2)(t).
This enables a very efficient evaluation as follows.
For each diagram class we take the \sg.(0)(s) diagram and specify an initial state $a = a'$, thereby fixing the final state $b=b'=a$ as well (\Fig{gl-4}, left most diagram).
We then determine all allowed intermediate states $c_1,c_2,c_3$ on the lower propagator.
For each such possible sequence of states $a,{c_i}$, the TMEs need to be evaluated only once per class.
We furthermore have to calculate only two energy dependent functions, one for \sg.(0)(s) and one for \sg.(2)(t), and then use
\begin{subequations}\begin{align}
  \textrm{\sg}.(1)(\textrm{s}) =  - \textrm{\sg}.(0)(\textrm{s}), \label{gl_s}\\
  \textrm{\sg}.(1)(\textrm{t}) =  - \textrm{\sg}.(2)(\textrm{t}),\label{gl_t}
\end{align}\end{subequations}
see \Fig{gl-4}. 
The energy dependent contributions times the TMEs can now simply be added to the respective kernel super matrix elements $(\hat{K}^{(4)}_\eff)^{aa}_{aa}$, $(\hat{K}^{(4)}_\eff)^{aa}_{c_ic_i}$, see \Fig{gl-4}.
Implementation of this scheme, utilizing the grouping, gain-loss relations and storage of/looping over non-zero TMEs only results in the speedup of about a total factor of 10-20 for the numerical calculations.

\subsection{Contributions of diagram classes to physical processes\label{physics}}
\begin{figure}
\includegraphics[width=0.82\columnwidth]{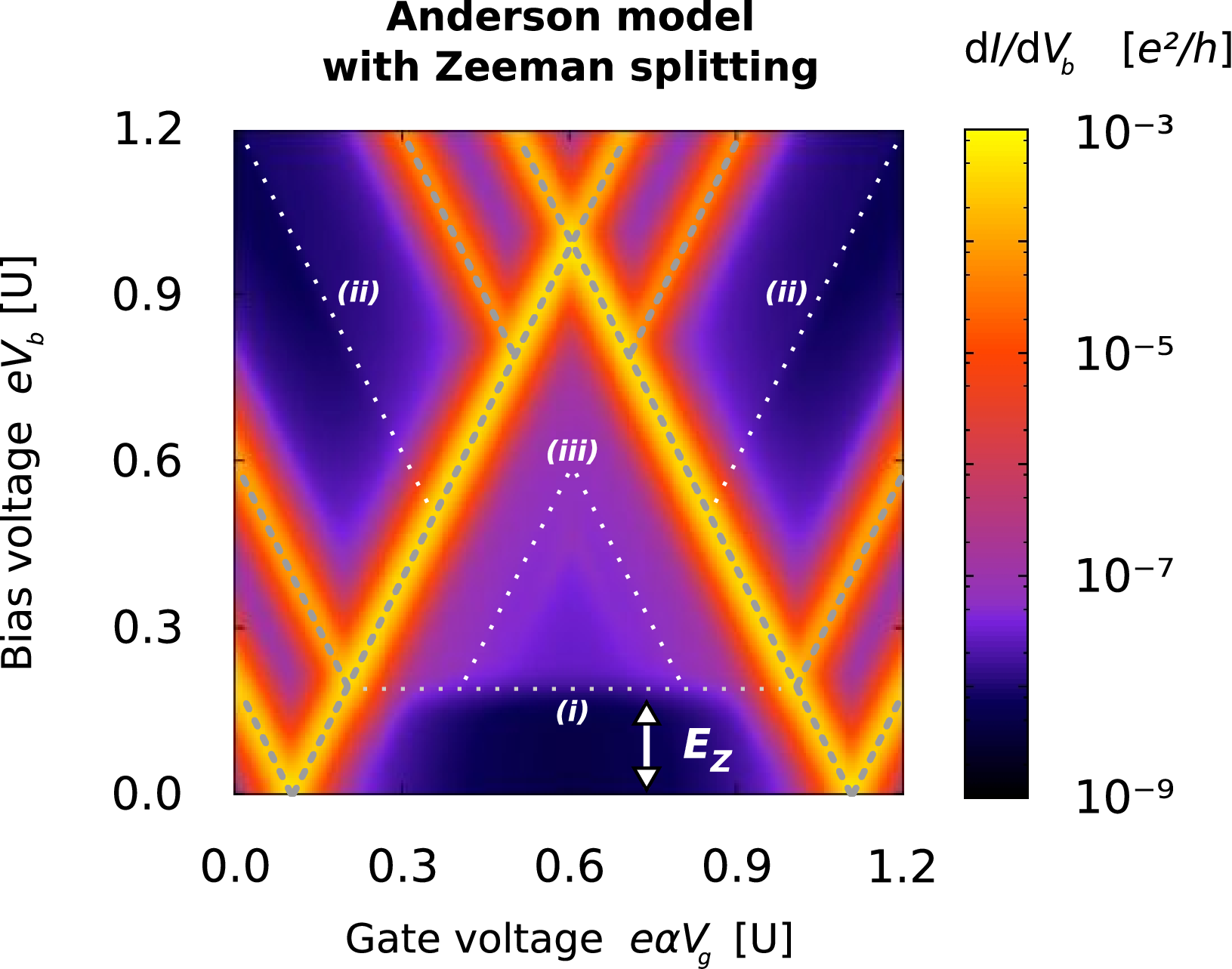}
\caption{Stability diagram ($\rmd I / \rmd V_\bias$ as a function of $V_\gate$ and $V_\bias$) for the Anderson impurity 
model in a magnetic field, see setup in \Fig{QD}(a).
The dot single-particle energy for spin-projection $\sigma = \; \downarrow / \uparrow$ 
is given by $\epsilon_\sigma = \epsilon_0 - e\alpha V_\gate \pm E_Z / 2$.
Here we set the Zeeman splitting to $E_Z = 0.2 U$, the level offset to $\epsilon_0 = 0.25 U$.
We used symmetric tunnel rates associated with the leads, $\Gamma_s = \Gamma_d = \Gamma = 0.0004\,k_BT / \hbar$
and set the thermal energy to $k_BT=0.01\,U$.
We assume equal capacitances associated with the source and drain tunnel junctions and apply the bias symmetrically.
The Coulomb  blockade regime, where the level is singly occupied, is the central triangular region within the shown gate range. The positions of the second order transport resonances are marked by dashed gray lines, the fourth order transport resonances (emphasized by white dotted lines) are (i) inelastic cotunneling, (ii) pair tunneling and (iii) cotunneling assisted sequential tunneling (weakly seen continuation of the Zeeman-lifted $|\downarrow\rangle\to|0/2\rangle$ excitation lines inside the Coulomb blockade region).
\label{SD-dIdV}}
\end{figure}
We now illustrate the importance of summing all diagrams of a given order in perturbation theory,
and the relative influence of the groups of diagrams.
This is relevant for understanding the impact of approximations, as well as the relation to the rates calculated in the \Tmat{} approach in~\Sct{T-mat}.
For this the simplest model of an interacting quantum dot, the Anderson impurity model, suffices.
This model describes a single level which can be populated by at most two (interacting) electrons of opposite spin:
\begin{equation*}
\hat{H}=\sum_{\sigma\in\{\uparrow,\downarrow\}} 
\epsilon_\sigma n_\sigma + U\hat{n}_\uparrow\hat{n}_\downarrow,
\end{equation*}
where $n_\sigma = d_\sigma^\dagger d_\sigma$ is the occupation number and $U$ the strength of the on-site Coulomb interaction
(excess energy required for double occupation).
The four many-body energy eigenstates are $|0\rangle$ (empty level), $|\sigma\rangle$ (singly occupied levels) and  
$|2\rangle$ (doubly occupied level), with energies $E_0 = 0$, $E_\sigma = \epsilon_\sigma$ and 
$E_2 = \sum_\sigma \epsilon_\sigma + U$. 
Under the influence of a magnetic field, the spin-degeneracy is lifted by the Zeeman splitting $E_Z= \epsilon_\uparrow - \epsilon_\downarrow$.
\par
\Fig{SD-dIdV} shows the corresponding stability diagram, i.e., conductance $\rmd I / \rmd V_\bias$ plotted as function of $V_\gate$ and $V_\bias$,
resulting from the full calculation including all second and fourth order contributions to the transport kernels.
We focus on gate voltages around the Coulomb blockade region where the dot is singly occupied.
In the chosen gate range the plot is left-right symmetric (particle-hole symmetry) and due to the additional
symmetry with respect to bias inversion (source-drain symmetry), only positive bias voltages are shown.
The ground-state to ground-state transitions, determining the edges of the singly occupied Coulomb blockade region, are due to the single-electron tunneling (SET) processes
$|\!\uparrow\rangle\to|0\rangle$ and $|\!\uparrow\rangle\to|2\rangle$. 
SET transitions involving the spin-excited state $|\!\downarrow\rangle$ appear as lines 
which are separated from the ground state transition lines $|0\rangle\to|\!\uparrow\rangle$ and $|2\rangle\to|\!\uparrow\rangle$ by the Zeeman energy $E_Z$. 

If the kernels were only calculated up to second order, only the above mentioned transport resonances would be seen in the 
stability diagram. Fourth order processes give rise to three additional types of transport resonances, see the numeration in \Fig{SD-dIdV}:

{(i)}
The horizontal step inside the Coulomb blockade region corresponds to the onset of inelastic cotunneling~\cite{Lambe68,DeFranceschi01}:
when the bias voltage exceeds the spin-splitting, $eV_\bias > E_Z$, a coherent tunnel process can take place which transfers an
electron from the source to the drain electrode, leaving the dot in the excited $|\!\downarrow\rangle$-state. 
This process involves only \emph{virtual} occupation of the energetically forbidden states $|2\rangle$ and $|0\rangle$ and is therefore 
only algebraically suppressed by the energy of these states. However, as it involves two coherent tunnel processes, it is proportional 
to the fourth power of the tunnel Hamiltonian $H_T$. Since the charge on the dot is the same before and after the cotunneling 
process, the resonance position is independent of $V_\gate$.

{(ii)}
There are additional steps in the differential conductance inside the SET regime, which have the same gate dependence as the SET resonances
(color change along lines ending at the upper figure corners).
These \emph{pair tunneling} resonances~\cite{Leijnse09b} correspond to direct transitions between the states 
$|0\rangle$ and $|2\rangle$, involving coherent tunneling of an electron pair onto / out off the dot.
Note that this becomes energetically allowed at a lower bias voltage than the sequential addition / removal of two electrons
($|0\rangle \to|\sigma \rangle \to |2\rangle$ / $|2\rangle \to|\sigma \rangle \to |0\rangle$).

{(iii)}
Finally, there are also gate-dependent peaks inside the Coulomb blockade regime,
the so-called cotunneling assisted sequential tunneling (CAST or CO-SET) resonances~\cite{Golovach04,Schleser05}. These correspond to SET transitions, where the \emph{initial} state is the excited $|\!\downarrow\rangle$-state. In second order only the ground state is populated inside the Coulomb blockade regime, such that this transition cannot take place. In fourth order, however, the excited state can be populated due to a preceding inelastic cotunneling process, and therefore resonance shows up above the inelastic cotunneling threshold inside the Coulomb blockade regime.

In addition to these resonance effects, fourth order terms both broaden and shift the SET resonances and give rise to a finite conductance background due to \emph{elastic} cotunneling (same as inelastic cotunneling explained above, but with initial and final states identical or of the same energy).\\

\begin{figure}[t]
\includegraphics[width=0.96\columnwidth]{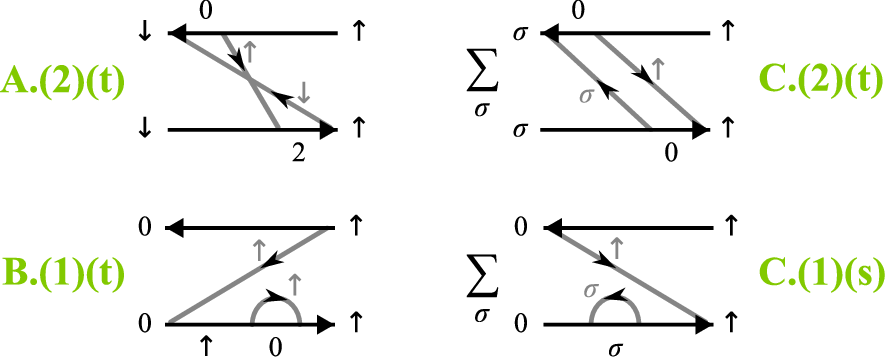}
\caption{Examples of diagrams with initial states $a=a'=\left|\uparrow\right\rangle$.
We determine the charge numbers of the allowed intermediate and the final states, using that at a vertex where a contraction line ends / starts the charge number changes by $\pm1$.
Similarly, the spin projection changes by $\pm\sigma/2$ at this vertex,
 where $\sigma/2$ is the electrode spin-index of the contraction.
Using this restriction, we find that the group A.(2) can only contribute to inelastic cotunneling ($b=b'=\left|\downarrow\right\rangle$),
 while C.(2) also allows the elastic process ($b=b'=\left|\uparrow\right\rangle$).
Groups B.(1) and C.(1) yield broadening and level renormalization of sequential tunneling processes, like $\left|\uparrow\right\rangle\to\left|0\right\rangle$:
B.(1) accounts for the possibility of a charge fluctuation in the initial state ($\left|\uparrow\right\rangle\to\left|0\right\rangle\to\left|\uparrow\right\rangle$), C.(1) for a charge fluctuation in the final state ($\left|0\right\rangle\to\left|\sigma\right\rangle\to\left|0\right\rangle$).
}
\label{statexample}
\end{figure}
\begin{figure*}
\includegraphics[width=1.94\columnwidth]{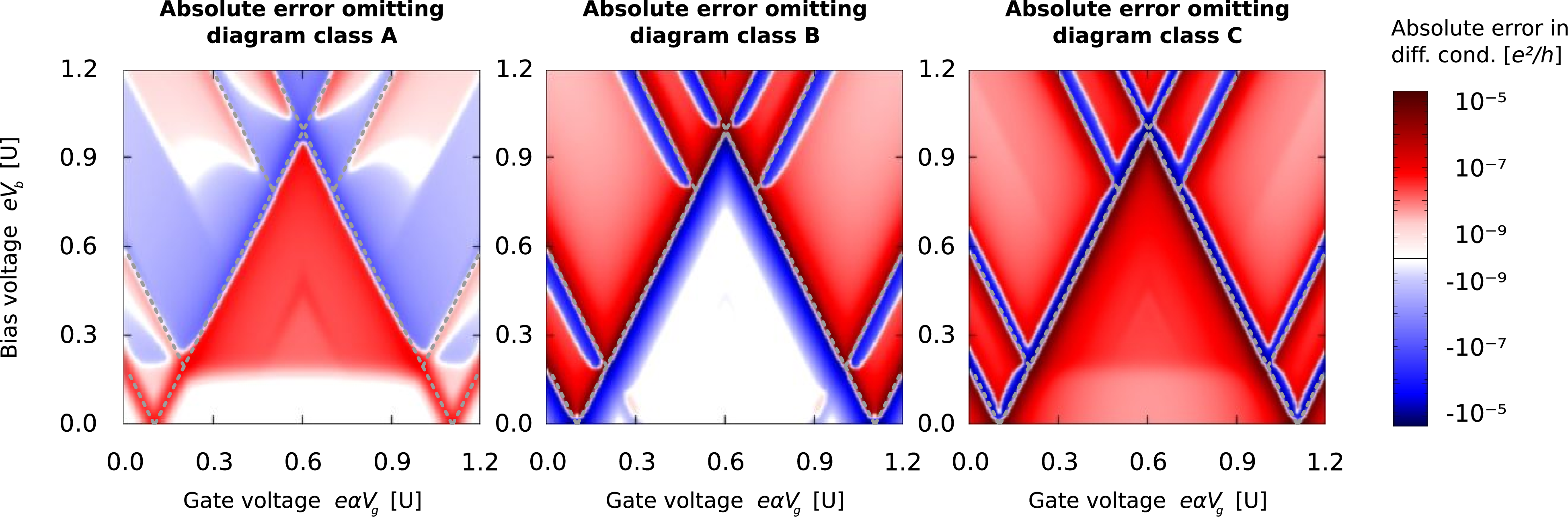}
\caption{Absolute errors occurring when neglecting contributions of a certain diagram class. Because for class B and C, the corrections along the resonance lines (marked by gray dotted lines) exceed the ones in between by orders of magnitude, the color scales are chosen logarithmic both in the positive and in the negative regime, i.e. ${}^{10}\text{log}|(\rmd I/\rmd V_\bias) - (\rmd I/\rmd V_\bias)_\text{approx}|$. Hereby, white color indicates that the contribution from the specific diagram class lies below the threshold for the logarithmic red / blue color scale applied for positive / negative error.
}\label{omit}
\end{figure*}

We now analyze to which extent the diagram groups in \Fig{diagr-tmat} contribute to specific rates for the physical processes mentioned above.
In \Fig{statexample} we illustrate for the initial state $|\uparrow\rangle$ how the selection rules for charge- and spin-projection at each vertex restrict the intermediate and the final states.
In general, 
 the allowed charge number $N_b$ of a final state $b$ for a given initial state $a$ with charge $N_a$
is readily found for each diagram group G.$(x)$ by assigning specific directions to the contraction lines and using \emph{charge selection rules only}.
Indicating these charge numbers in the kernel by $(K^{(4)}_\eff)_{N_b,N_a}$ we obtain  symbolically
\begin{subequations}\begin{align}
  (K^{(4)}_\eff)_{N_a,N_a}     \sim &  \mathrm{A}.(0)  + \mathrm{B}.(0) + \mathrm{C}.(0) +  \nonumber \\
                                  &  \mathrm{A}.(2)  + \mathrm{B}.(2) + \mathrm{C}.(2),   \label{cot}\\
  (K^{(4)}_\eff)_{N_a\pm1,N_a}  \sim &  \mathrm{A}.(1)  + \mathrm{B}.(1) + \mathrm{C}.(1),  \label{set}\\
  (K^{(4)}_\eff)_{N_a \pm2,N_a} \sim &  \mathrm{A}.(2)           + \mathrm{C}.(2).  \label{pt}
\end{align}\end{subequations}
From the restricted change in the charge numbers it is clear that \eq{cot} describes cotunneling, \eq{set} corrections to SET and \eq{pt} pair-tunneling.
We note a subtlety for \eq{cot}: if in addition to the energies, the initial and final \emph{states} are also equal ($a=b$), the rate must comply with the sum-rule \eq{sumrule}:
\begin{align}
    (K^{(4)}_\eff)^{aa}_{aa} = - \sum_{b \neq a}    (K^{(4)}_\eff)^{bb}_{aa}.
\end{align}
This means that \emph{elastic} cotunneling rates cannot be separated from the ``loss'' contributions which enforce the sum-rule.\\

To gain more insight into the physics incorporated in the different diagram groups,
we now selectively leave out contributions and calculate the resulting error in the current.
By pairwise neglecting horizontal neighbors in \Fig{diagr-tmat}, i.e., gain-loss partners (cf.,
\Sct{gain-loss}), the  sum-rule~(\ref{sumrule}) is conserved. However, the diagrammatic grouping
reveals that consistency is only guaranteed by neglecting entire classes of contributions.
This is illustrated by considering the contributions to the pair-tunneling rate \eq{pt}.
Assume that one considers neglecting the A.$(2)$(t) contribution in \eq{pt}. Then, to preserve the
sum-rule, we drop in \eq{set} the contribution of the A.(1)(t) subgroup. However,
  this occurs only in the combination A.$(1)=$A.$(1)$(t)$+$A.$(1)$(s) which contains
  physically necessary~\cite{koller_thesis} partial cancellations between the two terms:
  there are pair-tunneling contributions which should not influence the SET rate \eq{set}.
  We therefore drop together with A.$(1)$(t) also the A.$(1)$(s) term, and due to the sum-rule correspondingly in \eq{cot} the A.(0)(s) term. 
  To keep consistency, we thus exclude \emph{all} diagrams
  G$.(2)$(t), G$.(1)$(t), G$.(1)$(s) and G$.(0)$(s) of a certain class G.

In \Fig{omit} we illustrate the impact of the above for the Anderson model. Going from left to right we plot
the absolute error in the differential conductance $\mathrm{d}I/\mathrm{d}V_\bias$ resulting from the neglect of either diagram class A, B or C as a whole. Here blue (red) color indicates that \textit{inclusion} of the specific diagram class reduces (enhances) the differential conductance.
Lines along which the color changes from red to blue indicates an incorrect position of the resonance.
The occurrence of extended regions with uniform color
(constant differential error) additionally indicate that the current is wrong at all voltages above this region. 

The left panel in \Fig{omit} reveals that, for the Anderson model, class A does not have any influence below the inelastic cotunneling threshold. The reason is that by their structure the irreducible diagrams of the group A.$(2)$(t) necessarily involve a spin-flip as shown in~\Fig{statexample}) and therefore cannot contribute to the elastic cotunneling process $|\!\uparrow\rangle\to|\!\uparrow\rangle$.
Class A gives no major correction along the SET resonance lines as compared to classes B and C below.
However, inside the SET regime the increase of the conductance above the  pair tunneling resonance shows a large deviation when the class A contributions to pair tunneling are neglected.
Finally, the error made in the inelastic cotunneling also affects the relaxation of the spin excited state in CAST processes, as evidenced by the CAST resonance lines showing up in the error plot.

The situation is completely different when neglecting class B diagrams, as shown in the center panel of \Fig{omit}.
Deep inside the Coulomb diamond, they do not give any contribution, which is related to their topological structure.~\cite{koller_thesis}
Instead, class B yields significant contributions to the SET current level, as exhibited by the uniform positive (red) background in the SET regime. Moreover, the resonance positions are affected as well:
along each entire resonance line, pronounced ``shadows'' occur.
The negative (blue) correction below the resonance and the positive correction (red) above signal a shift of the onset of the current, i.e., a level renormalization~\cite{Koenig98}.
More precisely, class B diagrams can be related to \textit{level renormalization of the initial state} in a SET process.
In SET processes, the dot goes from state $a$ to state $b$, e.g. by addition of an electron, as represented by the left diagram in \Fig{gl-2}. Diagrams from group B.(1) have the same structure, except for an intermediate charge fluctuation (``bubble'') of the initial state $a$, see e.g. B.(1)(t) in \Fig{statexample} with  $a=a'=|\uparrow\rangle,\,b=b'=|0\rangle$.
The lowering of the energy of the initial state shifts the resonance positions of processes where electron tunnels  out (in)  to lower (higher) gate voltages, in agreement with the result in the center panel of \Fig{omit}.
See Ref.\cite{koller_thesis} for a more detailed discussion.
The pair tunneling is not affected since there are no B diagrams with two lines connecting upper and lower
parts of the contour, as required for two-electron transfer.

In a similar way, class C diagrams account for \textit{final state level renormalization}:
the color ``shadows'' along the resonance lines in the right panel of \Fig{omit} are practically inverted with respect to the result for class B in the center panel, indicating an opposite level shift.
Indeed, diagrams from group C.(1) contribute to tunneling from state $a$ to $b$ with an intermediate charge fluctuation in the final state $b$, see e.g. C.(1)(s) in \Fig{statexample} with $a=a'=|\uparrow\rangle,\,b=b'=|0\rangle$.
Additionally, class C contributes to inelastic cotunneling and completely mediates the elastic cotunneling process $|\uparrow\rangle\to|\uparrow\rangle$, see C.(2)(t) in \Fig{statexample}.
The error at the onset of pair tunneling, to which class C contributes, seems less pronounced than for class A. In fact the contributions are equal~\cite{Leijnse09b}
but this is masked by the uniform positive correction to SET processes which is larger for class C.

\section{Relation to the T-matrix approach\label{T-mat}}
Since the first studies of higher-order transport processes (see~\cite{Averin92}),
master equations have been used with transition rates calculated from a generalized form of Fermi's golden rule~\cite{Bruus} in many works studying transport beyond leading order in the tunnel 
coupling~\cite{Turek02, Golovach04, Paaske04Kondo, Koch04a, Misiorny06a, Elste06}.
This scattering or \Tmat{} approach seems to be very similar to the GME approaches discussed so far, and in this section, we clarify the connection.
For this the grouping of the kernel contributions discussed in the previous sections is of crucial importance.
In particular, it reveals the precise origin of the divergences occurring in the \Tmat{} rates
and allows us to \emph{derive} the correct regularization of these divergences,
 which differs from the ad-hoc regularization employed throughout the literature.
\subsection{Stationary state equation}
The idea of the \Tmat{} approach is to apply many-body scattering theory in the form of a generalized version of Fermi's golden rule to describe stationary quantum transport~\cite{Bruus}. One calculates the time evolution of the occupation probabilities of a state $\ket{\tilde{a}}$ of the \emph{total} system from the transition amplitudes
\begin{equation}
\left<\tilde{b}\left|\right.\tilde{a}(t)\right>=\left<\tilde{b}\left|\,\mathcal{T} \rme^{-\frac{\rmi}{\hbar}\int^{t}_{t_0}\rmd\tau\,H^I_T(\tau)\,}\right|\tilde{a}\right>\label{ttrate}
\end{equation}
where $\mathcal{T}$ denotes the time-ordering operator.
In \Eq{ttrate} it is assumed that at time $t_0$, when the interaction was switched on, the total system was in a direct product state
 $\ket{\tilde{a}}=\ket{a}\ket{k}$ of lead ($\ket{k}$) and quantum dot ($\ket{a}$) states. The leads are assumed to be individually at equilibrium at that time.
This exactly corresponds to the assumptions of the GME approach discussed in \Sct{easy}.
As a consequence of the interaction, the state $\ket{\tilde{a}}$ evolves  into $\ket{\tilde{a}(t)}$ for $t>t_0$, which is no longer of product form
and has an overlap with states $\ket{\tilde{b}} \neq \ket{\tilde{a}}$. The corresponding transition rate is calculated from the amplitudes via
\begin{align}
\gamma_{\tilde{a}\to\tilde{b}}=&\frac{\rmd}{\rmd t}\left|\left<\tilde{b}\left|\right.\tilde{a}(t)\right>\right|^2 \nonumber \\
=&\Re\left\{\left(\frac{\rmd}{\rmd t}\left<\tilde{b}\left|\right.\tilde{a}(t)\right>\right)\left<\tilde{a}(t)\left|\right.\tilde{b}\right>\right\}
\, .
\label{gammarate}
\end{align}
Usually, in the \Tmat{} approach only occupation probabilities are taken into account,
corresponding to diagonal elements of the RDM.
In general, this can be insufficient, as coherences between secular states (degenerate on the scale of $\Gamma$) play an important role for various models --\,e.g. in the case of (pseudo) spin polarization~\cite{Braun04set,Donarini06}.
Here we compare the effective rates determining the occupation probabilities in the \Tmat{} and \GME{} approach.
Averaging the transition rate (\ref{gammarate}), with $\ket{\tilde{a}} =
\ket{a}\ket{k} \to \ket{\tilde{b}} = \ket{b}\ket{k'}$, over the initial ($\ket{k}$) and final states ($\ket{k'}$) of the electrodes with the initial grand-canonical probabilities [cf.~\Eq{rholeads}] we obtain transition rates 
\begin{equation}\label{rare}
\Gamma_{a \to b} =
\sum_{kk'}
\gamma_{(a k) \to (bk')} \bra{k} \rho_\mathrm{R} \ket{k}
\end{equation}
for the time-evolution equation of the RDM,
\begin{equation}\label{re}\dot{\rho}_{bb}(t)=\sum_{a\neq b}\left[ \Gamma_{a\to b}(t,t_0)\,\rho_{aa}(t_0)-\Gamma_{b\to a}(t,t_0)\,\rho_{bb}(t_0) \right]
,
\end{equation}
see~\cite{Timm08} for a derivation.
The rates describe the probability for a transition to the state $|b\rangle$ at time $t$, given that the system was prepared in state $|a\rangle$ at time $t_0$~\cite{Timm08}.
Clearly, in the long time limit $t-t_0 \to \infty$ this is not an equation for the stationary state.
Still, the key step in the formulation of the \Tmat{} approach is that one replaces $\rho_{aa}(t_0)$ on the right-hand-side of \Eq{re} by the stationary occupancies $\rho_{aa}$, and sets $\dot{\rho}_{bb}(t)=0$. Then the resulting equation is solved for the occupancies, with the rates evaluated in the stationary limit $t_0 \to -\infty$.
In contrast, in the kinetic \noEq{equation}{rdm-compact}, the density matrix elements on the right-hand-side of the equation are not taken at the initial time $t_0$, but at times $\tau>t_0$ where the system has already approached the steady state.
We now first show that, as a direct result of the above ad-hoc replacement,
the \Tmat{} rates calculated to beyond second order in the tunneling $H_T$
are divergent in the zero frequency limit, i.e., $z \rightarrow 0$.

\subsection{Divergence of the stationary \Tmat{} kernel and its proper regularization}
There are two ways to calculate the rates, starting from the expansion of the time-ordered exponential in~\Eq{ttrate}:
\begin{multline}\left<\tilde{b}\left|\right.\tilde{a}(t)\right>=\Bigl<\tilde{b}\Bigr|\,1-\frac{\rmi}{\hbar}\int^{t}_{t_0} \rmd\tau \, H^I_T(\tau) \\ + 
\left(\frac{\rmi}{\hbar}\right)^2\int^{t}_{t_0}\rmd\tau\,H^I_T(\tau)\int^{\tau}_{t_0}\rmd\tau_1 \, H^I_T(\tau_1) - \cdots\,\Bigl|\tilde{a}\Bigr>.\label{tttime}\end{multline}
To make contact with the literature, we first follow the standard route by first performing the time-integrations, and obtain from \Eq{tttime} in the stationary limit $t_0 \to -\infty$
\begin{align}
\nonumber\left|\left<\tilde{b}\left|\right.\tilde{a}(t)\right>\right|&=\lim_{\eta\to0}\left|\frac{e^{\eta t}}{E_{\tilde{b}}-E_{\tilde{a}}+\rmi\eta}\left<\tilde{b}\left|T\right|\tilde{a}\right>\right|
,
\end{align}
such that $\gamma_{\tilde{a}\to\tilde{b}}$ becomes independent of $t$ as expected for the stationary state.
This gives the well-known generalization of the Golden-Rule rate\cite{Bruus}
\begin{align}
\gamma_{\tilde{a}\to\tilde{b}}&=2\pi\delta(E_{\tilde{b}}-E_{\tilde{a}})\left|\left<\tilde{b}\left|T\right|\tilde{a}\right>\right|^2,\label{tmat-rate}
\end{align}
where the transition amplitude involves the T-matrix $T$, instead of the interaction $H_T$. The T-matrix is defined by a Dyson-like equation
\[T=H_T+H_T\frac{1}{E_{\tilde{a}}-H_0+\rmi\eta}T,\] which can be truncated at the desired order.
For comparison with the GME approaches, it is more convenient to alternatively calculate $\gamma_{\tilde{a}\to\tilde{b}}$ to fourth order, postponing the time integrations. Setting $t_3=t$ in \Eq{ttrate}
\begin{multline*}\gamma_{\tilde{a}\to\tilde{b}}
=\Bigl<\tilde{b}\,\Bigr|\,\frac{1}{\rmi\hbar}H^I_T(t)\\\times\Bigl(1+\sum_{j=1}^3\left(\frac{1}{\rmi\hbar}\right)^{4-j}\prod_{k=j}^3\int^{t_k}_{t_0}\!\rmd t_{k-1}H^I_T(t_{k-1})\Bigr)\,\Bigl|\tilde{a}\Bigr>\\\times\Bigl<\tilde{b}\,\Bigr|\,1+\sum_{j=1}^3\left(\frac{1}{\rmi\hbar}\right)^{4-j}\prod_{k=j}^3\int^{t_k}_{t_0}\!\rmd t_{k-1}H^I_T(t_{k-1})\Bigr]\Bigl|\tilde{a}\Bigr>^{\dag}
\,.\end{multline*}
Since lead and quantum dot states at the initial time are not correlated we can now first trace out the leads.
One can express the \Tmat{} transition rates between states of the dot $a,b$ as the elements
\begin{align}
\Gamma_{a\to b}(t,t_0) =
\Bigl<b\Bigr|\,
\mathcal{K}^I_\text{TM}(t-t_0)
\left|a\left>\right<a\right|\Bigl|b\Bigr>
\label{rate4}
\end{align}
of the superoperator
\begin{multline}
\mathcal{K}^I_\text{TM}(t-\tau)=-\Tr_\leads \Lop(t)\Lop(\tau)\hrho_\leads\\
+\Tr_\leads\timeint{t}{\tau_2}{\tau_1}{\tau}\Lop(t)\Lop(\tau_2)\Lop(\tau_1)\Lop(\tau)\hrho_\leads.
\label{Tmatkernel}
\end{multline}
Explicitly, this follows by applying \Eq{merge} backwards. Alternatively, this equation results straightforwardly when formulating scattering theory~\cite{Mukamel82rev} in a Liouville or ``tetradic'' formalism~\cite{Fano63}.
\par
We can now make the connection to the kernel appearing in the \GME{}, which we discussed in \Sct{Nak-Zwa}.,
and compare the effective rate matrices for the occupancies.
Comparing with the \FQME{} superoperator in \Eq{\FQME-explicitkernel}, one finds that its second order part matches the one of \Eq{Tmatkernel} exactly.
Going to the Schr\"odinger picture and taking the Laplace transform [\Eq{kern-lapl}], and considering super-matrix elements between diagonal states we find
\begin{align}
 \left[{K}^{(2)}_\text{TM}(z)\right]^{aa}_{bb} =  \left[{K}^{(2)}_{ss}(z)\right]^{aa}_{bb}.
 \label{K2equal}
\end{align}
Thus, to the lowest order of perturbation theory, the \Tmat{} approach produces exactly the stationary GME equation with a kernel that is well behaved in the stationary limit $z \to i0$.

However, the fourth order part in \eq{Tmatkernel} is lacking the second term in \eq{\FQME-explicitkernel}
 which subtracts all the reducible parts among the fourth order contributions.
The physical origin of the appearance of the reducible correction term was traced clearly in the \FQME{}  derivation of the GME kernel in \Sct{Nak-Zwa} (and the NZ approach in App.~\ref{NZ}).
There the subtraction of reducible contributions, which is missing in \eq{rate4}, emerged by consistently eliminating $\hrho(t_0)$ in the second order term in favor of $\hrho(\tau)$.
(We note that in the RT approach this identification is harder to make, since one always deals with correctly regularized expressions from the start.)
This term thus accounts for the fact that at times $t>t_0$ the total system density matrix does not factorize anymore,
an effect which is however only important when going beyond the lowest order.
Indeed, one directly arrives at the \Tmat{} approach by ignoring this fact in the derivation of the \FQME{} approach.
Thus, in the \Tmat{} approach one effectively (but tacitly) makes the assumption that the dot and reservoir states
are statistically independent \emph{after} the interaction is switched on.
We now show that as a result of this assumption the rates in the \Tmat{} approach diverge in the stationary limit.
Writing out the relation between the fourth order parts of \Tmat{} and \FQME{} kernel, it reads in contrast to \Eq{K2equal}
\begin{align}
 \left[ {K}^{(4)}_{\text{TM}}(z)\right]^{aa}_{bb} 
=   \left[{K}^{(4)}_{ss}(z) +  {K}^{(4)}_\text{N}(z)  +  {K}^{(4)}_\text{S}(z)\right]^{aa}_{bb}.
\label{KTMz}
\end{align}
Here ${K}^{(4)}_\text{N}(z) + {K}^{(4)}_\text{S}(z)$ arises from the second reducible fourth order term in \eq{\FQME-explicitkernel},
which we decompose into two parts.
The first part contains only non-degenerate (non-secular) intermediate states,
\begin{align}
  {K}^{(4)}_\text{N}(z) & =  {K}^{(2)}_{sn}(z) \frac{i}{(-\mathcal{L}_{nn})} {K}^{(2)}_{ns}(z).
\end{align}
This is precisely the correction term to ${K}^{(4)}_{ss}$ in the effective master equation for the secular part of the density matrix (which here reduces to the diagonal part) arising from the non-diagonal coherences, cf.~\Eq{Keff}.
As explained in \Sct{nonsec} it must be included to obtain a systematic expansion of the effective transition rates between probabilities in powers of $\Gamma$.
Both kernels, ${K}^{(4)}_{ss}(z)$ and ${K}^{(4)}_\text{N}(z)$, are well-behaved for $z \to i0$ (see \Sct{group}).
The remaining term contains only intermediate states which are strictly degenerate in energy (strictly secular)
\begin{align}
 {K}^{(4)}_\text{S}(z) &=  {K}^{(2)}_{ss}(z) \frac{i}{z} {K}^{(2)}_{ss}(z)
 \, ,
 \label{KS}
\end{align}
and as a result diverges as $z^{-1}$ since ${K}^{(2)}_{ss}(z)$ is  well-behaved for $z \to i0$, cf.~\Eq{K2equal}.
Therefore the \Tmat{} rate ${K}^{(4)}_{\text{TM}}(z)$ diverges as $z^{-1}$ as well.
Rewriting \Eq{KTMz} we can express the effective \GME{} kernel \eq{Keff} determining the stationary occupation probabilities:
\begin{subequations}\begin{align}
  {K}^{(4)}_\eff (i0) & = {K}^{(4)}_{ss}(i0) + {K}^{(4)}_\text{N} (i0) \label{Keffgme}\\
                    &  =  \lim_{z \to i0}  \left( {K}^{(4)}_\text{TM}(z) - {K}^{(4)}_\text{S}(z) \right). \label{Kefftmat}
\end{align}\end{subequations}
This equation summarizes the central relation of \GME{} approaches to an \emph{automatically regularized \Tmat{} expression}.
The effective fourth order kernel for the probabilities is thus obtained either from \Eq{Keffgme} adding the non-secular reducible correction to the GME kernel (both finite) or from \Eq{Kefftmat} subtracting from the \Tmat{} kernel~\eq{Tmatkernel} the secular reducible correction and canceling the $z^{-1}$ divergences.
We have thereby precisely identified the correct regularizing term \eq{KS}
which should be used if one would like to keep on using a fourth order \Tmat,
expressed in terms of known second order \Tmat{} rate expressions [cf.~\Eq{K2equal}] and the frequency $z$.
We emphasize furthermore that the form \eq{Keffgme} makes explicit that the probabilities contain corrections from the non-secular coherences
whereas \eq{Kefftmat} and, in fact, the \Tmat{} approach itself, make no reference to non-diagonal density matrix elements.
\par
\Eq{Kefftmat} shows explicitly that the relevant kernel for transport and scattering problem are related.
General relations between irreducible kernels and Liouville \Tmat{} expressions were given first by Fano~\cite{Fano63}.
However, to our knowledge, the relation between the \Tmat{} rates and the non-secular corrections to the effective fourth order kernel, has not been addressed before.

\subsection{Regularization error for Anderson model}
A key point of the \Tmat{} approach as formulated in the literature is that instead of~\Eq{Kefftmat}
 one uses an ad-hoc regularization to cure the divergence
which was inadvertently introduced by the ad-hoc formulation of the stationary state equation.
The resulting \Tmat{} rates for occupations obtained in the literature show a striking similarity to the \GME{} expressions for the effective rates for the occupancies, \emph{including the non-secular contributions from the coherences}.
The reason for the similarity is also made explicit in \Eq{Kefftmat}.
However, we now explicitly show that the ad-hoc regularization significantly differs from the correct regularizing term \Eq{KS} subtracted in \Eq{Kefftmat}. To illustrate its importance analytically, we consider the simple Anderson model already studied in \Sct{physics}.
In applications of the \Tmat{} approach, typically not all fourth order contributions are included. Very commonly, corrections to SET, i.e., diagram groups A.(1), C.(1) and the complete class B, are dropped a priori. As this breaks the gain-loss-chain (see \Sct{physics}), the sum-rule is then enforced by hand, which makes the groups A.(0) and C.(0) redundant as well.

As we do \emph{not} wish to study here the errors arising from such additional approximations (see Ref.~\cite{Begemann10} for such a comparison), we focus on a cotunneling contribution which would be included in any type of \Tmat{} based calculation, namely the one from diagram group C.(2).$(t)$ (contributions from A.(2).$(t)$ drop out for infinite charging energy) to the elastic process $|\sigma\rangle\to|0\rangle\to|\sigma\rangle$. We compare the energy dependent parts of its analytical contribution to the kernel element ${K}^{\sigma\sigma}_{\sigma\sigma}$ for the \GME{}, respectively $\left({K}_\text{TM}\right)^{\sigma\sigma}_{\sigma\sigma}$ for the \Tmat{}.
In \Sct{group}, we determined the energy dependent part of the contribution of subgroup C.(2)(t) to be [\Eqs{expr1_triple}{expr1_C}]:
\[\mathrm{C}.(2)(\mathrm{t})\sim\left\{\begin{array}{rl}\left[{\delta_1}\,{\delta_2}\,({\delta_3-\delta_2+\rmi0})\right]^{-1}&\delta_2\neq\delta_1+\delta_3-\rmi 0,\\
\left[{\delta_1}\,{\delta_2}\,{\delta_3}\right]^{-1}&\delta_2=\delta_1+\delta_3-\rmi 0.
\end{array}\right.
\]
This function clearly distinguishes between the cases of secular and non-secular intermediate states, avoiding the inclusion of the divergent reducible term.
For our example, we determine $\delta_1,\delta_2,\delta_3$ diagrammatically using \Fig{deltacuts}.
For the kernel element ${K}^{\sigma\sigma}_{\sigma\sigma}$ in the Anderson model, the states are $a=a'=b=b'=\sigma$, $c=d=0$, where $\sigma\in\{\uparrow,\downarrow\}$:
\begin{eqnarray*}\delta_0&=&0\\ \delta_1&=&\rmi 0 +E_0-E_\sigma+\omega',\\\delta_2&=&\rmi 0 -\omega+\omega',\\\delta_3&=&\rmi 0 +E_\sigma-E_0-\omega.\end{eqnarray*}
The corresponding \Tmat{} expression misses the exclusion of the reducible diagrams and is always given by 
\begin{eqnarray*}\mathrm{C}^{\mathrm{TM}}(2)(\mathrm{t})&\sim&\left[{\delta_1}\,{\delta_2}\,({\delta_3-\delta_2}+\rmi 0)\right]^{-1}\\
&=&\frac{1}{\omega-\omega'-\rmi 0}\frac{1}{\left|\omega'+E_0-E_\sigma+\rmi 0\right|^2},\end{eqnarray*}
even when the secular condition $\delta_2+\rmi 0=\delta_1+\delta_3$ is fulfilled here.
In the final expression for the rate, $\mathrm{C}^{\mathrm{TM}}(x)(\mathrm{t})$ is multiplied by two Fermi functions and we integrate over their arguments, $\omega$ and $\omega'$ respectively.
Performing the $\omega'$ integral first, one is left with a divergent $\omega$ integral
due to the modulus-square factor in $\mathrm{C}^{\mathrm{TM}}(2)(\mathrm{t})$.
The standard way to regularize the \Tmat{} rates mentioned above now proceeds as 
follows~\cite{Turek02,Koch04a}.
For any function $F(\omega)$  which is well-behaved at $\omega=0$ one expands in the infinitesimal $i0$
\begin{align}
\nonumber\int\rmd\omega\frac{F(\omega)}{\omega^2-(\rmi0)^2}&=\int\rmd\omega\frac{F(0)}{\omega^2-(\rmi0)^2}+\int\rmd\omega\frac{F(\omega)-F(0)}{\omega^2-(\rmi0)^2}\\&=\frac{\pi}{\rmi0}F(0)+\int^\prime\rmd\omega\frac{F(\omega)-F(0)}{\omega^2-(\rmi0)^2}
,
\label{Tmat-regl}
\end{align} \noindent where $\int^\prime$ denotes a principal part integration. Exploiting $1=\frac{1}{1+e^\omega}+\frac{1}{1+e^{-\omega}}$ it can be shown that $\int^\prime\frac{\mathrm{d}\omega}{\omega^2-(i0)^2}\to0$.
Divergent contributions $\propto1/(\rmi0)$ are claimed to be due to sequential tunnel processes which are already included in other rates
and are ignored.
As shown above in general, however, the divergent term is due to neglecting the mixing of the lead and dot states, which is \emph{not} an effect of sequential tunneling (see~\Eq{K2equal}).
Moreover, the regularization procedure does in fact \textit{not} reproduce the regularization which is automatically included in the \GME{} approach.

\begin{figure}
\includegraphics[width=0.82\columnwidth]{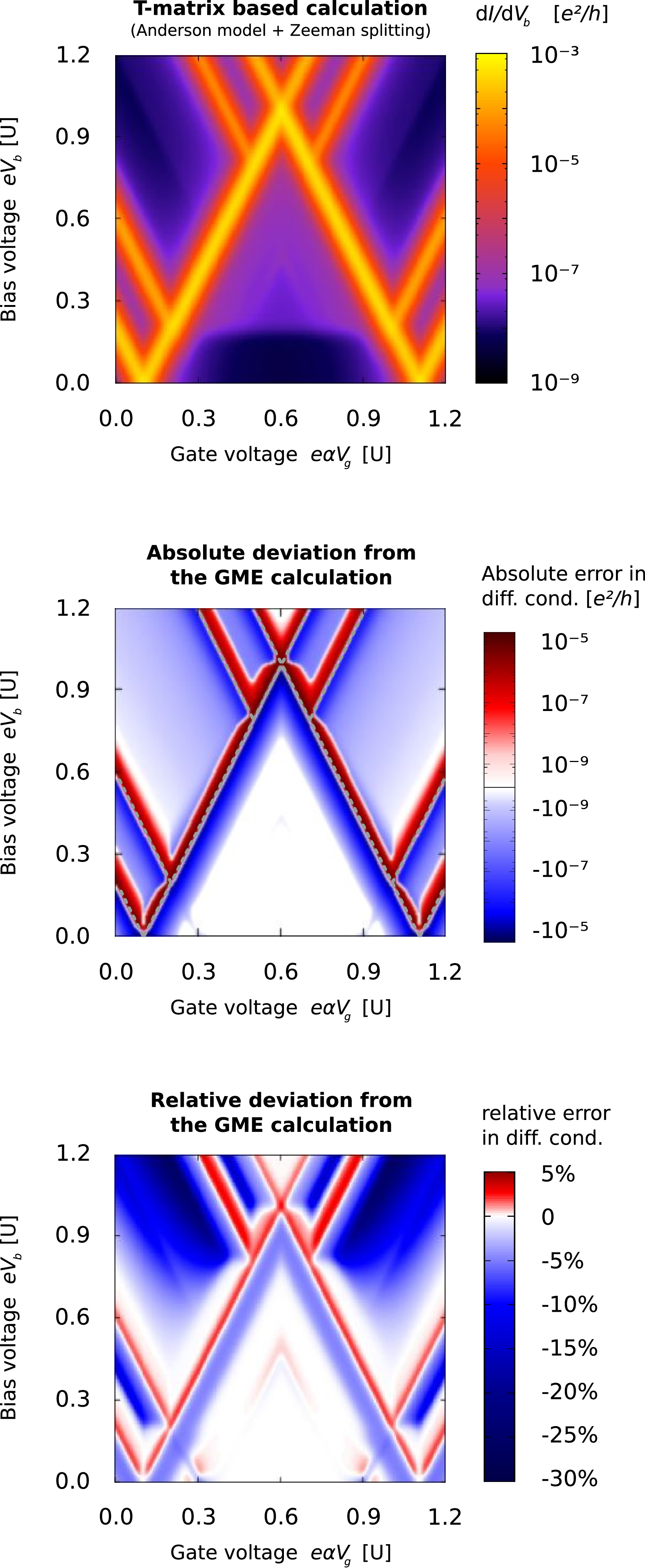}
\caption{Top: The stability diagram for a single-level quantum dot as in \Fig{SD-dIdV} calculated using the \Tmat{} approach.
Middle: Absolute deviation between the non-linear conductance calculated with the \GME{} and \Tmat{} approach, in logarithmic scale, i.e. ${}^{10}\text{log}|(\rmd I_\text{GME}/\rmd V_\bias) - (\rmd I_\text{TM}/\rmd V_\bias)|$, with color coding just as in \Fig{omit}.
Bottom: Relative deviation between the non-linear conductance calculated with the \GME{} and \Tmat{} approach, $\left[(\rmd I_\text{GME}/\rmd V_\bias)-(\rmd I_\text{TM}/\rmd V_\bias)\right] / (\rmd I_\text{GME}/\rmd V_\bias)$.
Although the upper panel seems similar to the GME result in \Fig{SD-dIdV}, indeed large relative errors exist in the vicinity of resonances and
throughout the regime where transport is not suppressed by Coulomb blockade,
in particular when electron pair-tunneling in the single-electron tunneling regime~\cite{Leijnse09b} is energetically allowed.
\label{dIdVTmat}}
\end{figure}

To illustrate this quantitatively and gain insight into which voltage regimes this matters, we have plotted in \Fig{dIdVTmat} (top panel) the differential conductance computed with the \Tmat{} approach, using the same model and parameters as in \Fig{SD-dIdV}.
To demonstrate that existing discrepancies are not simply healed by
including the typically neglected corrections to SET, we have employed here
the ``best-possible'' \Tmat{} kernel, taking into account all contributions
arising from \Eq{Tmatkernel}, just regularizing the occurring divergences
according to \Eq{Tmat-regl}. For diagram class B, such regularization leads
indeed to a complete omission when $\delta_2=i0$ (secular intermediate free
propagating states), which is the case for the Anderson model in our
example. Class A, containing only irreducible diagrams, requires no
regularization and is taken into account fully. Both the \GME{} kernel as
well as this ``best-possible'' \Tmat{} kernel satisfy the
probability-conservation sum-rule. The associated current kernels are
constructed as discussed in \Sct{easy}.

In the lower panel we show the relative deviation between the \Tmat{} and \GME{} results, blue (red) color indicating that the \GME{} differential conductance falls below (exceeds) the one obtained from the \Tmat{} approach.
Clearly, the agreement is good deep inside the Coulomb blockade region.
However, all resonance lines are dressed by a pronounced red (blue) shadow from above (below),
 indicating that the corrections to sequential tunneling (level renormalization, broadening) are not correctly taken into account in the \Tmat.
For this simple setup, the maximum of deviation encountered amounts to $30\%$ overestimation and $5\%$ underestimation of the correct result.
Given the parameters of the model, we have actually given with \Fig{dIdVTmat} the best result that can be possibly obtained within the \Tmat{} approach with the \Eq{Tmat-regl}.
This includes the effects of elastic and inelastic cotunneling, pair tunneling and single-electron level renormalization and broadening effects.
As mentioned before, often other simplifications are made, in addition to the above procedure employed, leading to further deviations.
For instance, quite commonly, only the fourth order cotunneling rates from the diagram groups A.(2) and C.(2) are taken into account.
The figure shows that the \Tmat{} approach basically only works in the ``deep Coulomb blockade regime'' where
SET \emph{and} CO-SET processes are suppressed.

\section{Summary\label{summary}}
In summary, we have studied the systematic calculation of the fourth order kernel within the generalized master equation approach for transport through quantum dots.
At present this is the only approach which can efficiently deal with strongly interacting systems with complex excitation spectra in the non-linear transport regime,
while accessing the regime of moderate tunnel coupling by a perturbative treatment of higher order tunnel processes.
The precise understanding of the calculation of the kernel determining the transport rates is therefore of great practical importance.
On the one hand, simplifications which speed up numerical calculations are crucial to allow more complex physics to be addressed.
On the other hand, the comparison of calculations with kernels evaluated using different methods is an urgent issue. We summarize our main achievements:

We have first shown the equivalence between the real-time diagrammatic approach (\RT{}) and the Bloch-Redfield quantum master equation (\FQME{}) (and the Nakajima-Zwanzig technique). This was done both formally and explicitly, by mapping irreducible \RT{} diagrams onto operator expressions in the \FQME{}.
In particular, we showed that in the \FQME{}, the well-behaved kernel from the diagrammatic approach is obtained as a sum of two parts: One part contains all (irreducible and reducible) terms of a certain order. The other part exactly cancels all the reducible terms, which diverge in the stationary (zero-frequency) limit.

Next we addressed the calculation of the density matrix using this kernel.
The commonly used secular approximation for the reduced density matrix was previously shown to break down when going to next-to-leading order in the tunneling~\cite{Leijnse08a}.
Despite this, an effective kernel determining the secular part of the reduced density matrix can be derived:
Here we showed that the non-secular corrections can effectively be accounted for with a secular kernel by including certain reducible diagrams with non-secular intermediate states.
By adding these to the standard irreducible diagrams a new structure is revealed.
All diagrams can be sorted into groups of diagrams with the same distribution of vertices over the Keldysh contour and the same contractions.
We showed that subgroups consisting of three diagrams can be summed analytically,
 thereby avoiding unnecessary integral evaluations.
We derived new diagram rules for evaluating an entire subgroup at once,
 arriving at an expression as simple as that for a single diagram. The physics behind diagrams and groups was illuminated by a study of specific contributions for a single level with finite Coulomb interaction and Zeeman splitting (single impurity Anderson model).

Furthermore, the summation of subgroups allows for a general comparison between the \GME{} and \Tmat{} approach (``generalized Golden Rule''):
We first showed formally that the \Tmat{} ``rate-kernel'' equals the correct fourth order kernel plus a divergent term.
Thereby we identified precisely the correct term needed to regularize the \Tmat{} rates.
We emphasize that the GME (BR or RT) kernels are well-behaved and finite by construction, the regularization is incorporated automatically and not put in ``by hand''.
We showed both numerically and analytically that the regularizations existing in the literature are incorrect
and lead to analytical deviations in the individual rates and, as a result, to pronounced errors in the calculated transport current if one is not deep inside the Coulomb blockade regime.
We illustrated this numerically for the example of the Anderson model in magnetic field.

Finally, we note that the key ideas we presented can be applied to
perturbation calculations beyond the first two leading orders for simple models
and other classes of problems which are formulated in the same way.
The latter include the calculation of noise~\cite{Thielmann05}, adiabatic time-dependent transport~\cite{Splettstoesser06} and the renormalization group extension of the GME approach~\cite{Schoeller09a}.

\acknowledgements{ Financial support under the DFG programs SFB689 and SPP1243 and the European Union under the FP7 STREP program SINGLE is acknowledged. We thank Herbert Schoeller,
Jens Paaske and Georg Begemann for valuable discussions.}
\appendix

\section{Nakajima-Zwanzig projection technique\label{NZ}}
The standard Nakajima-Zwanzig projection operator technique~\cite{Nakajima58,Zwanzig60} allows for a compact and concise derivation of an exact expression for the kernel, see e.g.~\cite{Breuer,Fick} and the references therein.
We briefly review the notation and the steps involved in the derivation of the \Kkernel{} in the time-domain
which is performed concisely in the interaction picture. In a similar fashion, one can derive the current kernel.
Using projectors $\P= \hrho_R \Tr_\leads$ and $\Q=\hat{1}-\P$
we decompose the total density matrix, $\hrho^{I}_\tot= \P\hrho^{I}_\tot+ \Q\hrho^{I}_\tot $,
and project the interaction-picture Liouville equation~(\ref{liouville-new}) for the full system:
\begin{align} 
  \P\dot{\hrho}^{I}_\tot  &= -\rmi\, \hat{\mathcal{P}} \Lop(t) \hat{\mathcal{Q}}\, \hrho^{I}_\tot
  ,
  \label{eq:Pdotrho}
  \\
  \Q\dot{\rho}^{I}_\tot  &= - \rmi\, \Q \Lop(t) \Q \hrho^{I}_\tot - \rmi\, \Q \Lop(t) \P \rho^{I}_\tot.
  \label{eq:Qdotrho}
\end{align}
Here, the crucial property $\P \Lop(t) \P =0$ was used, which is due to the fact that the tunnel Hamiltonian~(\ref{lt_ham}), 
and thus also $\Lop$~(\ref{lioudef}), contains exactly one lead operator: the trace must yield zero.
Next, the second equation is formally integrated using $\Q\hrho_\tot^{I}(t_0)= 0$ and treating the term with $\P\rho_\tot^{I}$ as a given inhomogeneous term:
\begin{align}
  \Q\hrho_\tot(t)  &=  -\rmi
  \int^t_{t_0}\! \rmd\tau_1\,
  \hat{\mathcal{T}} e^{-\rmi \int_{\tau_1}^{t}\! \rmd\tau \Q\Lop(\tau)\Q } \Q \Lop(\tau_1) \P \hrho_\tot (\tau_1).
  \label{eq:Qrho}
\end{align}
Here $\hat{\mathcal{T}}$ is the time-ordering superoperator.
Substitution into \Eq{eq:Pdotrho} gives the kinetic equation~(\ref{\FQME-compact}) with the formally exact kernel:
\begin{align}
  \hat{\mathcal{K}}^I(t,\tau) =  - \Tr_\leads \left(\hat{\mathcal{T}} \Lop(t) e^{-\rmi \Q \int^{t}_{\tau} \rmd\tau_1 \Lop(\tau_1) \Q} \Lop(\tau) \hrho_\leads\right),
  \label{eq:KIproj}
\end{align}
which transforms into the well-known result for the \Kkernel{} in the Schr\"odinger picture, \Eq{thekernel}.
The \Kkernel{}~(\ref{eq:KIproj}) contains the non-trivial evolution operator which can be expanded in the perturbation
$\Q \Lop(\tau_1)\Q$,
\begin{align}
  &e^{-\rmi \Q \int^{t}_{\tau} \rmd\tau_1 \Lop(\tau_1) \Q}
  =1 -\rmi
   \int^{t}_{\tau}\rmd\tau_1\ 
   \Q\Lop(\tau_1)\Q
   \nonumber \\
   &
   -
   \timeint{t}{\tau_2}{\tau_1}{\tau}
   \ \Q \Lop (\tau_2) \Q \Lop (\tau_1) \Q
   + \ldots\ .
\end{align}
Applying this we obtain the kernel to fourth order, which can be compared to \Eq{\FQME-explicitkernel} to confirm the equivalence to the \FQME{} approach. Inserting $\Q=\hat{1}-\P$ everywhere and using that $\P$ gives non-zero only when acting on an even number of $\Lop$ by Wick's theorem we obtain
\begin{align}
  & \hat{\mathcal{K}}^{I}(t,\tau) =
  -\Tr_\leads
  \Lop(t)\Lop(\tau) \hrho_\leads
  \nonumber\\
  &
  +\Tr_\leads 
  \timeint{t}{\tau_2}{\tau_1}{\tau}
  \Lop(t)  \Lop(\tau_2) (\hat{1}-\P) \Lop(\tau_1)  \Lop(\tau) \hrho_\leads.
  \label{eq:KIproj4}
\end{align}
The first fourth order term, involving the unit operator 1 in the middle, gives rise to all possible contractions from which the reducible ones are subtracted by the second term with $\P$ in the middle.
Alternatively, one may also first integrate \Eq{eq:Pdotrho} for $\P\hrho_\tot^{I}$ with initial condition
$\P\hrho_\tot^{I}(t_0)= \hrho_\tot^{I}(t_0)$:
\begin{align}
  \P\hrho^I_\tot(t)  &=  \hrho_\tot (t_0)
  -\rmi \int^{t}_{t_0}\rmd\tau_2
                      \P \Lop(\tau_2) \Q\hrho^I_\tot (\tau_2).\label{eq:prho}
\end{align}
Substitution of \Eq{eq:Qrho} into the right hand side of \Eq{eq:prho} and taking the trace gives a Dyson-type integro-differential equation for the reduced density operator $\hrho^I = \Tr_\leads \P \hrho^I_\tot$.
The equivalent equation for the propagator defined by
$
\hrho^I(t) =\hPi^I(t,t_0)\hrho^I(t_0)
$
reads
\begin{align}
  \hPi^I(t,t_0) = \hat{1} +
  \timeint{t}{\tau_2}{\tau_1}{t_0}
  \hat{\mathcal{K}}^I(\tau_2,\tau_1) \hPi^I(\tau_1,t_0),
\end{align}
with the kernel given by \Eq{eq:KIproj}, revealing total equivalence to \Eq{pi-rt} as obtained in the RT approach.

\section{Time- and frequency space calculation of the kernels}
In this appendix we present the details of the calculation of the kernel \Eq{thekernel} starting from the interaction-picture expansion used in the \FQME{} approach in \Fig{diagr-table}.
We explicitly obtain a result which was mentioned in \Sct{diagr-repr} and used as a starting point in \Sct{group} of the main text: contributions to the kernel represented by diagrams differing only by relative time-ordering of vertices on different parts of the contour, deviate only in the time-dependent function or its Laplace transform~\eq{expr1}.
In \App{app:example} we first discuss an example calculation of second and fourth order kernel contributions for the Anderson model, starting from the expressions given in \Fig{diagr-table},
for readers not familiar with either the \RT{} or \FQME{} technique.
In \App{app:diagr_rules} we summarize the general diagram rules, i.e., how to set up the \Kkernel{} in terms of diagrams and how to directly read off the final expression for the contribution from each diagram.

\subsection{Examples\label{app:example}}
We consider the Anderson impurity model with finite Coulomb interaction introduced in~\Sct{physics},
 characterized by the four many-body states $|0\rangle,\ |\!\!\uparrow\rangle,\ |\!\!\downarrow\rangle,\ |2\rangle$, corresponding respectively to zero, one spin-up, one spin-down or two spin-paired electrons on the dot. We demonstrate the technique by evaluating only the kernel element $(K)^{22}_{22} = \bra{2} \left[ K \ket{2} \bra{2} \right] \ket{2}$,
 which contains in second order the two ``loss'' rates for processes $|2\rangle\to|\sigma\rangle$. In fourth order it includes besides ``loss'' rates also the elastic cotunneling $|2\rangle\to|2\rangle$.\\

For the calculations we will need the following, generally valid transformation for kernel elements from time space to Laplace space,
\begin{multline}\label{KtimeKfreq}
\left(K(z)\right)^{aa'}_{bb'}=\int_0^\infty e^{iz\tau'}\,\Bigl\langle b\Bigr|\, e^{-\frac{i}{\hbar}Ht}  \\\times\mathcal{K}(\tau')[e^{\frac{i}{\hbar}H(t-\tau')}\left|a\right\rangle\left\langle a'\right|e^{-\frac{i}{\hbar}H(t-\tau')}]e^{\frac{i}{\hbar}Ht}\Bigr| b'\Bigr\rangle,
\end{multline}
where $\tau'=t-\tau$. As compared to \Eq{kern-lapl}, additional exponentials arise because of the fact that matrix elements with respect to the states of the RDM --\, which transforms according to \Eq{ia-to-schr}\, -- have been taken.

\subsubsection*{2nd order}
From \Fig{diagr-table} we can infer which diagrams contribute by using the charge-selection rule: at each vertex the charge changes by one. For the Anderson model, starting from $\ket{2}$ only the intermediate state $\ket{\sigma}$ is possible, with either spin $\sigma = \uparrow, \downarrow$.
The zero-frequency contribution reads
\begin{center}
\begin{tabular}{lr}
\raisebox{0.52cm}{$\left(\hat{{K}}^{(2)}\right)^{22}_{22}=$}
&\includegraphics[width=0.58\columnwidth]{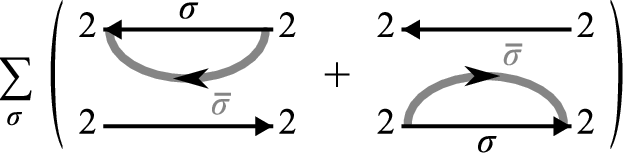}
\end{tabular}\end{center}
This is a sum of two complex conjugate expressions, so we only need to evaluate e.g. the last complex expression.
From \Fig{diagr-table} we obtain the time-dependent interaction picture expression,
which we transform to the Schr\"odinger picture and Laplace transform with respect to the time-interval $\tau'$ spanned by the diagram and send $z \to i0$\\
\begin{tabular}{l@{\hspace{-1.5cm}}l}
\raisebox{-0.4cm}{\includegraphics[width=0.18\columnwidth]{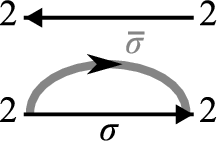}}
&\begin{minipage}{3cm}\begin{multline*} 
\qquad\quad\quad=-\lim_{z\to\rmi0^+}\int_0^\infty\!\!\rmd\tau'\ \rme^{iz\tau'}\left\langle\hat{C}^{-}_0\hat{C}^{+}_3\right\rangle
\\
\times\left\langle 2\right|\,\rme^{-\frac{\rmi}{\hbar}\hat{H}\tau'}\,\left| 2\right\rangle\cdot\left\langle 2\right|\,\rme^{-\frac{\rmi}{\hbar}\hat{H}(t-\tau')}\hat{D}^{+}_0\left|\sigma\right\rangle
\left\langle\sigma\right|\,\hat{D}^{-}_3\,\rme^{\frac{\rmi}{\hbar}\hat{H}t}\,\left| 2\right\rangle.
\label{example2}\end{multline*}
\end{minipage}\\&
\end{tabular}\\
Here the expressions left and right of the $\cdot$ correspond to the upper and lower contour, respectively.
The exponentials containing $\hat{H}$ arise from transforming the operator expression from the interaction to the Schr\"odinger picture.
Next we transform the occurring operators $D$ and $C$ to the Schr\"odinger picture, according to \Eqs{transD}{transC}. With the use of \Eqs{substitution}{substitution2} as well as \Eqs{tme_manybody1}{tme_manybody} we find\\
\begin{tabular}{l@{\hspace{-0.3cm}}l}
\raisebox{0.2cm}{\includegraphics[width=0.18\columnwidth]{app_exmpl_2nd_spec.eps}}
&\begin{minipage}{3cm}\begin{multline*}=-\hbar^{-2}\sum_l\int_0^\infty\!\!\rmd\tau'\ \int\!\rmd\omega\ \tilde{\rho}_{l\bar{\sigl}}(\omega)\\\times\,\rme^{\frac{\rmi}{\hbar}\left(E_\sigma-E_2+\omega+\rmi 0\right)\tau'}f_{l}^{-}(\omega)\Tp_{l\bar{\sigl}}(2,\sigma)\Tm_{l\bar{\sigl}}(\sigma,2)
\,\\
=-\frac{\rmi}{\hbar}
\sum_l\int\!\rmd\omega\ \frac{f_{l}^{-}(\omega)\tilde{\rho}_{l\bar{\sigl}}(\omega)}{\omega+E_\sigma-E_2+\rmi 0}\left|\Tp_{l\bar{\sigl}}(2,\sigma)\right|^2,
\end{multline*}
\end{minipage}\\&
\end{tabular}\\
where $\bar{\sigma} = - \sigma$.
Here  $f^{+}_l(\omega)\equiv f( (\omega-\mu_l)/(k_BT))=(e^{(\omega-\mu_l)/(k_BT)}+1)^{-1}$ is the Fermi function of lead $l$ with temperature $T$, $f^-_l(\omega)\equiv1-f_l(\omega)$, and
$\tilde{\rho}_{l\sigl}(\omega)=\sum_q\delta(\epsilon_{l\sigma q}-\omega)$, as it occurs in \Eq{Gammal}, is the (possibly spin-dependent) density of states in lead $l$. Notice that for our simple Anderson model we have no $q$ dependence of the single particle tunneling amplitudes, and thus there is no $q$ dependence of the TMEs here.
Furthermore, the time $t$ cancels out in the exponentials after performing the $\tau'$ integral.
This example illustrates the form of the time-evolution factor of any 2nd order contribution:
\begin{equation*}
\frac{1}{\Delta_0} = \frac{1}{\delta_1}
\end{equation*} 
where $\Delta_0$ is the sum of energies occurring in the argument of the exponential which contains the time $\tau'$.

\subsubsection*{4th order}
For the Anderson model with non-magnetic electrodes, selection rules
cause the fourth order non-secular corrections to vanish, i.e., $K_{ns}=0$ in \Eq{diagramcorrections}.
Therefore only irreducible contributions remain.
These selection rules can be used furthermore, in addition to the charge selection rule,
to determine the allowed intermediate states on the diagrams below. For the example kernel element the charge number does not change, and therefore all contributing diagrams have an even number of vertices on each contour (i.e., groups \sg.(0) and \sg.(2) in \Fig{diagr-tmat}). Notice further that there are no diagrams from group A.(2), because these would involve an intermediate charge state with three electrons on the dot.
\begin{center}\begin{tabular}{lr}\\
\raisebox{3.5cm}{$\left(\hat{{K}}^{(4)}\right)^{22}_{22}=$}
&\includegraphics[width=0.72\columnwidth]{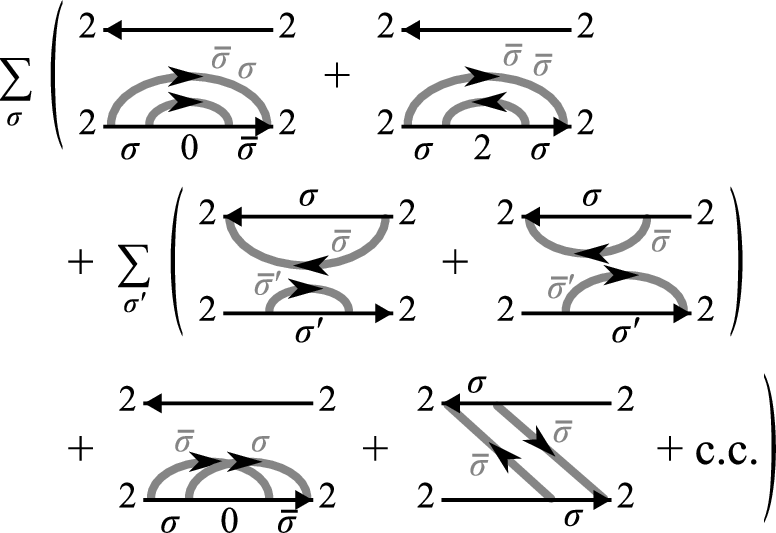}
\end{tabular}\end{center}
We calculate the last shown diagram, starting from the expression given in \Fig{diagr-table} and introduce the time-distance of vertex $i$ to the final time $\tau'_i:=t-\tau_i$, and perform the same steps as in the second order example:\begin{widetext}
\begin{tabular}{@{\hspace{-0.4cm}}l@{\hspace{-0.1cm}}l}
\raisebox{1.15cm}{\includegraphics[width=0.09\columnwidth]{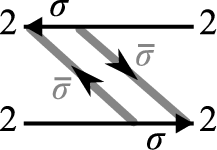}}
&\begin{minipage}{3cm}\begin{eqnarray*}
&=&-\lim_{z\to\rmi0^+}\int_0^\infty\!\!\rmd\tau'\ \rme^{iz\tau'}\int^{t}_{t-\tau'}\!\!\rmd\tau_1\int^{t}_{\tau_1}\rmd{\tau_2}\left\langle\hat{C}^-_0\hat{C}^+_2\right\rangle\left\langle\hat{C}^+_1\hat{C}^-_3\right\rangle\\
&&\times\left\langle 2\left|\,\rme^{-\frac{\rmi}{\hbar}\hat{H}t}\,\hat{D}^+_3\right|\sigma\right\rangle\left\langle\sigma\left|\hat{D}^-_2\,\rme^{\frac{\rmi}{\hbar}\hat{H}(t-\tau')}\,\right|2\right\rangle\cdot\left\langle 2\left|\,\rme^{-\frac{\rmi}{\hbar}\hat{H}(t-\tau')}\,\hat{D}^+_0\right|\sigma\right\rangle\left\langle\sigma\left|\hat{D}^-_1\,\rme^{\frac{\rmi}{\hbar}\hat{H}t}\,\right|2\right\rangle\\
&=&\hbar^{-4}
\sum_{ll'}\int_0^\infty\!\!\!\!\rmd\tau'\!\int_{0}^{\tau'}\!\!\!\!\rmd\tau'_1\!\int_{0}^{\tau'_1}\!\!\!\!\rmd\tau'_2\int\!\rmd\omega\!\int\!\rmd\omega'\tilde{\rho}_{l\bar{\sigl}}(\omega)\tilde{\rho}_{l'\bar{\sigl}}(\omega')\,\rme^{\frac{\rmi}{\hbar}(-\omega+E_2-E_\sigma)\tau'_1}\rme^{\frac{\rmi}{\hbar}(-\omega'+E_2-E_\sigma)\tau'_2}\rme^{\frac{\rmi}{\hbar}(\omega'-E_2+E_\sigma+\rmi 0)\tau'}\\
&&\times f_{l}^{+}(\omega)\,f^{-}_{l'}(\omega')\,\Tp_{l\bar{\sigl}}(2,\sigma)\Tm_{l'\bar{\sigl}}(\sigma,2)\Tp_{l'\bar{\sigl}}(2,\sigma)\Tm_{l\bar{\sigl}}(\sigma,2)\\
&=&-\frac{\rmi}{\hbar}
\sum_{ll'}\int\!\rmd\omega\int\!\rmd\omega'\,\left|\Tp_{l\bar{\sigl}}(2,\sigma)\right|^2\left|\Tp_{l'\bar{\sigl}}(2,\sigma)\right|^2\frac{1}{-\omega+\omega'+\rmi 0}\frac{f_{l}^{+}(\omega)\tilde{\rho}_{l\bar{\sigl}}(\omega)}{-\omega+E_2-E_\sigma+\rmi 0}\,\frac{f^{-}_{l'}(\omega')\tilde{\rho}_{l'\bar{\sigl}}(\omega')}{\omega'-E_2+E_\sigma+\rmi 0}.
\end{eqnarray*}\end{minipage}\\&
\end{tabular}\end{widetext}
In this example all contracted electrode operators have inverted time-order (corresponding to the earliest vertex being on the lower contour):
as a result the $\omega$ contraction gets the $f^{+}_l$ function in contrast to the contraction in the second order example.
We note that the structure of the Laplace transformed time-evolution factor appearing in the fourth order contributions has the general form:
\[
\frac{1}{\Delta_0}\frac{1}{\Delta_0+\Delta_1}\frac{1}{\Delta_0+\Delta_1+\Delta_2}
=\frac{1}{\delta_1\delta_2\delta_3},
\]
where $\Delta_{0/1/2}$ denote the arguments of the exponentials containing the times $\tau'/\tau'_1/\tau'_2$, respectively.

\subsection{Diagram rules for zero-frequency kernel\label{app:diagr_rules}}
We now give the rules by which one can directly write down the diagrammatic representation of the \emph{effective} \Kkernel{} and afterwards simply read off from each of these single (ungrouped) diagrams the resulting analytical contribution to $K^{(n)}_\eff(z)$, \Eq{Keff}, in the zero-frequency limit $z \to i0$. We illustrate how this applies to the examples of the previous section. Notice that modifications of these ``traditional'' diagram rules in order to account for a whole subgroup of diagrams were presented in \Sct{triple}.

\subsubsection*{The kernel in diagrammatic representation}
The rules for \emph{drawing} all diagrams representing the effective kernel of even order $n=2,4,..$ are as follows:
\begin{itemize}
\item
  Draw all distributions of $n$ vertices over the two contours, vertex $k$ being at time $\tau_k$, $k=0,\ldots,n-1$.
  Vertices $n-1$ and $0$ are at the boundaries of the diagram at times $\tau_{n-1}=t$ and $\tau_0=\tau$, respectively.
\item
  For each distribution, contract all $n$ vertices in $n/2$ pairs, denoting each contraction by a directed line.
  Each resulting diagram represents a distinct contribution to the \emph{effective} \Kkernel{}.
  Note that  all irreducible \emph{and} reducible contractions need to be included,
  where irreducible diagrams are those which can nowhere be vertically cut without cutting a contraction line.
\item
  To each contraction $j$ {($1\leq j\leq n/2)$},
  assign an energy $\omega_j$, as well as lead and spin indices $l_j,\,\sigma_j$, respectively.
\item
  On each contour, assign to each segment between two vertices a many body state of the quantum dot.
\end{itemize}
For the previous examples in second and fourth order we thus obtain:
\begin{center}
\begin{tabular}{rl}
\raisebox{0.9cm}{\framebox{$2nd$}$\quad\Rightarrow\qquad$}&
\includegraphics[width=2.6cm]{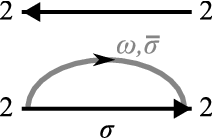}
\\\\
\raisebox{0.9cm}{\framebox{$4th$}$\quad\Rightarrow\qquad$}&
\includegraphics[width=2.6cm]{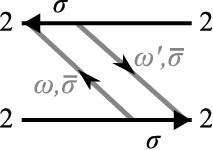}
\end{tabular}
\end{center}

\subsubsection*{Translating a diagram}
The rules for \emph{translating}  a diagram into an analytical expression can be divided into rules for determining three factors.
\begin{enumerate}
\item
  For each contraction $j$ write a Fermi distribution function and spin-dependent density of states
  \begin{align}
    f^{\pm}_{l_j}(\omega_{j}) \tilde{\rho}_{l_j\sigma_j}(\omega_j)
  \end{align}
  where $+$ ($-$) is chosen if the contraction line agrees (disagrees) with the \emph{contour direction}.
  For the vertex at which the contraction starts / ends write a many-body tunnel matrix element
  \[ T^{\pm}_{l_j\sigma_j}(b,a)
  ,
  \]
  where  $a$ and $b$ are the states before and after the vertex, respectively, following the direction of the \emph{contour} (not of time).
  
  \begin{tabular}{lll}\framebox{$2nd$}&$\Rightarrow$&$f_{l}^{-}(\omega)\tilde{\rho}_{l\bar{\sigl}}(\omega)\Tp_{l\bar{\sigl}}(2,\sigma)\Tm_{l\bar{\sigl}}(\sigma,2)$\\\\
    \framebox{$4th$}&$\Rightarrow$&$f_{l}^{+}(\omega)\,f^{-}_{l'}(\omega')\tilde{\rho}_{l\bar{\sigl}}(\omega)\tilde{\rho}_{l'\bar{\sigl}}(\omega')$\\
    &&$\times\Tp_{l\bar{\sigl}}(2,\sigma)\Tm_{l'\bar{\sigl}}(\sigma,2)\Tp_{l'\bar{\sigl}}(2,\sigma)\Tm_{l\bar{\sigl}}(\sigma,2)$
  \end{tabular}
\item
  Determine the time evolution factor
  \begin{align}
    \prod_{k=0}^{n-1} \frac{1}{\delta_{k}}
  \end{align}
  by drawing through each segment of the diagram between consecutive times $\tau_k$ and $\tau_{k+1}$ ($0 \leq k \leq n-2$) a vertical cut (see \Figs{gl-2}{gl-4} in the main part of the text).
  Obtain the denominator $\delta_k$ by adding / subtracting the energies of the dot states on the contour and the energies of the contractions depending on whether they hit the vertical cut from the left or from the right.
  Also add to $\delta_k$ the frequency $z=\rmi0$.
  
  \begin{tabular}{lll}\framebox{$2nd$}&$\Rightarrow$&$\delta_1+\rmi0=E_\sigma+\omega-E_2+\rmi0$\\\\
\framebox{$4th$}&$\Rightarrow$&$\delta_3+\rmi0=E_2-\omega+E_\sigma+\rmi0,$\\
                &             &$\delta_2+\rmi0=E_2-\omega+\omega'-E_2+\rmi0,$\\
                &             &$\delta_1+\rmi0=E_\sigma+\omega'-E_2+\rmi0.$
  \end{tabular}

(Note that $i0$ is not a convergence factor put in by hand, but naturally arises from the Laplace transform, reflecting the correct analytic behavior of the kernel.)
\item
  For the diagram as a whole determine the phase
  \begin{align}
    -\frac{i}{\hbar}(-1)^{n_\text{c} + n_\text{l}}
  \end{align}
  by counting the number of crossing contraction lines $n_\text{c}$, and the number of vertices on the lower contour $n_\text{l}$.
  
  \begin{tabular}{l}
    \framebox{$2nd$}\framebox{$4th$}\quad{$\Rightarrow -\frac{i}{\hbar}(-1)^{0+2}=-\frac{i}{\hbar}$}
  \end{tabular}\\
\end{enumerate}
Finally one multiplies the three factors, integrates over all frequencies $\omega_j$ and sums over all spin values $\sigma_j$ and electrodes $l_j$.
Notice that also all possibilities for intermediate quantum dot many body states on the contour have to be summed over.
The diagram rules, as formulated above, provide the key insight needed in the main text:
since diagrams within a \emph{group}, as defined in \Sct{group},
are related by moving their vertices around on each part of the contour, only the factor arising from the time-evolution is different.
The ordering of the vertices, the direction of the contractions relative to the \emph{contour}, as well as the number of crossing lines and vertices are all preserved under this operation.

The derivation of the rules in the form presented above can be found in~\cite{Schoeller97hab}.
Because of the importance of the time-evolution factors we comment on the relation between $\delta_k$ as defined by
\begin{align}
  \delta_k = \sum_{l=0}^{k+1} \Delta_{l}
\end{align}
and the value obtained by diagram rule 2.
This is easily seen once one notices the diagrammatic meaning of $\Delta_l$ (see~\cite{Schoeller97hab}, p. 95):
it equals the sum of energies of all lines going into the vertex $l$ minus those of the lines going out,
i.e., the dot energies of the in- and out-going lines of the contour and the energy of the contraction which starts / ends at vertex $l$.
Summing contributions from all vertices from earlier times $\tau_l \leq \tau_k$, the energies of all contractions which have started and ended thus cancel out, leaving the difference of energies of contractions running backward and forward with respect to \emph{time}.
In the sum, the dot energies of subsequent vertices on the same part of the contour cancel out, leaving only the difference between the upper and lower contour dot energy between vertex $k$ and $k-1$.
Summing all  $\Delta_k$ from both contours, one gets
$\sum_{k=0}^{n-1}\Delta_k=0$
This condition ensures that finally no exponential containing $t$ is left,
reflecting that the kernel, see \Eq{rdm-compact}, depends merely on the time difference $\tau'=t-\tau$. A more explicit version of this proof can be found in Ref.\cite{koller_thesis}.

\section{Diagram grouping\label{app:grouping}}
In this appendix we prove the statement made in \Sct{group}: freely integrating over the intermediate 
time $\tau_2$ in the representative diagrams is equivalent to summing the three diagrams within the corresponding 
subgroup, provided that the initial states (earliest times) are degenerate.
Additionally, we show how to easily calculate the partial summation of one or two irreducible diagrams 
in a subgroup in the case where secular diagrams have to be excluded.
This exclusion is automatically obtained in our transport theory and is an important
result of the paper: it prevents the divergences which plague the \Tmat{} approach from appearing.
We present the derivation both in time- and frequency-representation which each have their distinct advantages.

As explained in the main text, we want to express the sum of diagrams in a subgroup in terms of the contribution of a representative diagram (topmost diagram in each subgroup in \Fig{diagr-tmat}).
The standard diagram rules express this contribution as the product of propagators over time-intervals of length 
$-\tilde{\tau}_k=\tau_k-\tau_{k-1}$, see \Eq{eq:one_diagram}:
\begin{align}\nonumber
\mathrm{G}.(x)(\mathrm{t})\sim&
  \int_0^\infty\!\rmd\tilde{\tau}_1 \rmd\tilde{\tau}_2 \rmd\tilde{\tau}_3\ e^{
    -\frac{i}{\hbar} \left(
      \delta_3\tilde{\tau}_3+\delta_2 \tilde{\tau}_2+\delta_1 \tilde{\tau}_1
      \right)
  }\\
  \equiv
  &\int_0^\infty\!\rmd\tilde{\tau}_1 \rmd\tilde{\tau}_2 \rmd\tilde{\tau}_3\
  g(\tau_3,\tau_2,\tau_1,\tau_0).
  \label{eq:g-time-ref}
\end{align}
Here $\delta_k$ is the sum of energies of the backward minus the forward moving contour parts and contraction lines
in the segment of the diagram between times $\tau_k$ and $\tau_{k-1}$, see \Fig{deltacuts},
including also the Laplace variable $z \to \rmi 0$.
The simplifications only work for the zero-frequency Laplace transform of the kernel
which is all that is needed for the stationary state. 
\par
\begin{figure}
\includegraphics[width=\columnwidth]{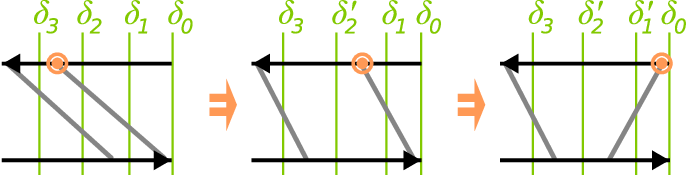}
\caption{From the representative diagram, all members of the triple group can be constructed, 
exemplified here by subgroup C.($2$)(t).\label{grouptrans}
}
\end{figure}

We now allow the next-to-last vertex (at time $\tau_2$) in the representative diagrams 
to move to earlier times (to the right in the diagram), see \Fig{grouptrans}.
Thereby we generate the other diagrams in the subgroup.
In the first step, the vertex at $\tau_2$ is permuted with the one at time $\tau_1$ and thereby the energy difference $\delta_2$ of the segment bounded by these vertices is changed to a different value: $\delta_2 \rightarrow {\delta}'_2$.
Permuting the vertex (now at time $\tau_1$) in the second step with the vertex at $\tau_0$, the energy difference $\delta_1$ changes as well:
$\delta_1 \rightarrow {\delta}'_1$.
From the diagram rules it follows that for each such permutation the following relation holds:
\begin{align}
 \delta_{n}+ {\delta}'_{n} =  {\delta}'_{n+1} + \delta_{n-1}, 
\end{align}
i.e. the average of the old and new value (left) equals the average of energy differences of the two adjacent segments (right).
Note that on the left the energy has already been modified to ${\delta}'_{n+1}$ by the preceding 
permutations (with the exception of the latest one $\delta_3$).
This is the key relation allowing the summation of diagrams within each subgroup.
We now apply this to the three diagrams in \Fig{grouptrans} and obtain
\begin{align}
 \delta_{2}+ {\delta}'_{2} & =   {\delta}_{3} + \delta_{1},
 \label{eq:delta2sum}
 \\
 \delta_{1}+ {\delta}'_{1} & =   {\delta}'_{2} + \delta_{0}.
 \label{eq:delta1sum}
\end{align}
Combining these we obtain the relation
\begin{align}
 \delta_{2}+ {\delta}'_{1} & =   {\delta}_{3} + \delta_{0}.
 \label{eq:delta12sum}
\end{align}
Here $\delta_0  = E_a - E_{a'}$ denotes the energy difference of the initial states, i.e. \emph{outside} the diagram,
and does \emph{not} include the Laplace variable $z=i0$ as the $\delta_k,k=1,2,3$ do.
Using this, 
 the contributions from the generated diagrams can now be expressed in the representative function, with the time-arguments of the corresponding vertices permuted:
\begin{align}
  e^{
    -\frac{i}{\hbar} \left(
      \delta_3\tilde{\tau}_3+{\delta}'_2 \tilde{\tau}_2+\delta_1 \tilde{\tau}_1
      \right)   
   }
   &=  g(\tau_3,\underset{\uparrow}{\tau_1},\underset{\uparrow}{\tau_2},\tau_0),  \label{eq:g-time-ref2}
   \\
  e^{
    -\frac{i}{\hbar} \left(
      \delta_3\tilde{\tau}_3+{\delta}'_2 \tilde{\tau}_2+{\delta}'_1 \tilde{\tau}_1
      \right)
   }
   &=  g(\tau_3,\underset{\uparrow}{\tau_0},\tau_2,\underset{\uparrow}{\tau_1}).
   \label{eq:g-time-ref3}
\end{align}
The arrows indicate how, by permuting time-arguments, \eq{eq:g-time-ref2} is obtained from  \eq{eq:g-time-ref},
and \eq{eq:g-time-ref3} from \eq{eq:g-time-ref2}.
Although this result may seem obvious, one should note that the last equation only holds under the condition $\delta_0 =  0$.
Thus, the diagram obtained by permuting the latest vertex with one of the earlier ones can only be expressed in the representative diagram if the initial states are degenerate.
Time-ordered integration of the sum of all the diagrams in the subgroup is now seen to be equivalent to decoupled time-integrations on opposite parts of the Keldysh contour of the single representative diagram, neglecting the remaining diagrams from the group:
\begin{widetext}
\begin{center}\begin{multline}
$\sg.($x$)(t)$ \sim 
\underset{\tau_3 > \tau_2 > \tau_1 > \tau_0 > -\infty}{\int \rmd \tau_2\, \rmd \tau_1\, \rmd \tau_0 }
\bigl[g(\tau_3,\tau_2,\tau_1,\tau_0)+g(\tau_3,\tau_1,\tau_2,\tau_0)+g(\tau_3,\tau_0,\tau_2,\tau_1)\bigr]
=
\timeint{\tau_3}{\tau_1}{\tau_0}{-\infty}\ \ \underset{\tau_3 > \tau_2 > -\infty}{\int\rmd\tau_2}\ 
g(\tau_3,\tau_2,\tau_1,\tau_0)
\label{merge}
\end{multline}\end{center}
\end{widetext}
Inserting the form of $g$ from \Eq{eq:g-time-ref}, and changing variables to time-intervals on the \emph{separate contours},
$ \hat{\tau}_2  =\tau_3-\tau_1$, $\hat{\tau}_1=\tau_1-\tau_0$ (forward) and
${\hat{\tau}'}_1=\tau_3-\tau_2$ (backward) the integrals decouple as usual.
We obtain the result in the main part of the paper:
\begin{align}
\text{\sg.(}x\text{)(t)} \sim 
\frac{1}{ (i0+\delta_{3}-\delta_2){\delta }_{2}{\delta }_{1} } & & \text{non-secular}
 \label{eq:sum}
\end{align}
Note that the $i0$ has to be supplied ``by hand'' explicitly since it formally cancels in the difference $\delta_{3}-\delta_2$.
Below we show that this does not alter the value of the integral
and that the sign automatically follows from the correctly regularized terms which are being summed.

For diagram class A this completes the derivation. However, diagram classes B and C contain
reducible diagrams, which diverge if one would allow for secular intermediate states.
In this case, one thus has to perform a partial sum of the irreducible diagrams in the subgroup only.
Note that in our formalism the exclusion of these cases is automatically enforced, i.e.,
we \textit{do not} exclude them ``by hand'' based on the mere inconvenience of divergent terms.

Although the explicit result for classes B and C can be obtained in the same way as above, we now show that here
 the frequency space representation has definite advantages.

We first derive \Eq{eq:sum} again in frequency-space by directly summing the Laplace
transforms of the propagators [left hand side of \Eq{eq:g-time-ref} and \Eqs{eq:g-time-ref2}{eq:g-time-ref3}]
\begin{align}
 \text{\sg.(}x\text{)(t)} \sim
 \frac{1}{ {\delta}_{3}{\delta }_{2}{\delta }_{1} } +
 \frac{1}{ {\delta}_{3}{\delta}'_{2}{\delta}_{1} } +
 \frac{1}{ {\delta}_{3}{\delta}'_{2}{\delta}'_{1} } \nonumber \\
  =
 \frac{1}{ {\delta}_{3}{\delta }_{2}{\delta }_{1} } +
 \frac{1}{ {\delta}_{3}{\delta}'_{1}{\delta }_{1} }
  =
 \frac{1}{ {\delta}'_{1}{\delta}_{2}{\delta}_{1} } & & \text{non-secular}
 \label{eq:sum-freq}
\end{align}
We first performed the partial sum over the last two terms using \noEq{relation}{eq:delta1sum} 
and assuming $\delta_0 = 0$.
Adding the last term and using \noEq{relation}{eq:delta2sum} we obtain the full sum.
Expressing ${\delta}'_1$ in the energy differences of the representative diagram $\delta_3,\delta_2,\delta_1$ using \Eq{eq:delta12sum} we again obtain \Eq{eq:sum}.
However, in addition we have treated the cases of secular intermediate states as well.
For the B.(1) and B.(2) subgroups, the representative diagram is itself reducible and the secular case arises for $\delta_2 =i0$.
If it is excluded, we keep only the partial sum in the second line of \Eq{eq:sum-freq}:
\begin{align}
 \text{B.(}x\text{)(t)} \sim \frac{1}{ {\delta}_{3}^2 {\delta}_{1} } &  & \text{secular}
\end{align}
Note that here the $i0$ does not need to be written out explicitly.
In contrast, for the C.(1) and C.(2) subgroups, the representative diagram is the only irreducible one which has to be kept: inspecting the reducible diagrams we see that the secular case arises for ${\delta}'_2 =\delta_3-\delta_2+\delta_1 =i0$ and we can eliminate one parameter. Keeping $\delta_3$ we obtain
\begin{align}
 \text{C.(}x\text{)(t)} \sim \frac{1}{ \delta_3 (\delta_{3}+\delta_{1}) \delta_{1} } &  & \text{secular}
\end{align}
We note that also here $i0$ needs not be written out explicitly since the $2i0$ in $\delta_{3}+\delta_{1}$ are sufficient to guarantee the correct analytic behavior as function of the frequencies.
We now shortly comment on this point as well on the $i0$ explicitly added in \Eq{eq:sum}.

Above we have calculated the time-integrals of the time-evolution factors in the diagrams only.
The resulting expressions, multiplied by the statistical factors (Fermi-functions), still need to be integrated over the frequencies of the contractions which are included in the energy differences $\delta_k$.
These integrals are exactly those of the standard perturbation theory and can be found in, e.g., Refs.~\cite{Leijnse08a,koller_thesis}.
However, in the above (partial) summations over subgroup diagrams, we have at several instances used that we can replace $i0$ by $2i0$. This does not alter these integrals.
The integrand possess a countable number of poles and decays sufficiently fast for the residue theorem to apply. Closing the integration contour in the upper half of the complex plane, we enclose the very same poles whether we take $i0$ or $2i0$.
The results of the frequency integrations are thus unaltered by the (partial) summation of subgroup diagrams as performed above.

Finally, we note that we have demonstrated all key ideas required for application to other problems.
For example, calculations of other stationary transport quantities (e.g. noise, adiabatic time-dependent transport) involve the same type of Keldysh diagrams \cite{Thielmann05,Splettstoesser06} and may be simplified exploiting the above technique.
Higher order perturbation calculations may also come within reach.
Importantly, the relative computational gain allowed by the simplifications reported here increases with the order of perturbation theory.
Consider, for example, \emph{sixth order} in $H_T$ (neglecting the technicalities of excluding secular cases for simplicity).
With 6 vertices to be distributed over the two parts of the contour, taking into account all possible contractions and directions of fermion lines, there are $7680$ diagrams.
The equivalent to \Fig{diagr-tmat} thus contains $320$ irreducible and $160$ reducible diagrams. 
However, one can identify 15 diagram classes comprising groups with $x\in\{0, 1, 2, 3\}$ vertices on the upper contour, 
containing 1, 1+5, 5+10, 10 diagrams respectively. The $x=1$ groups split into one stand-alone diagram (the gain-loss partner of the $x=0$ diagram) plus a subgroup of five diagrams. The $x=2$ groups comprise in turn the subgroup of (gain-loss partners of those) five diagrams plus a subgroup of ten diagrams (gain-loss partners of the $x=3$ diagrams). Summing the diagrams in each subgroup and using that only one subgroup from each pair of gain-loss partners needs to be evaluated (cf.~\Sct{gain-loss}), our grouping method reduces the number of expressions to be calculated by a factor of $8$.

\bibliography{bv10893}
\bibliographystyle{apsrev}
\end{document}